\documentclass[11pt]{article}
\usepackage{amsmath}
\usepackage{amssymb}
\usepackage{bm}
\usepackage{bbold}
\usepackage{hyperref}
\usepackage{subcaption}
\usepackage{graphicx}
\usepackage[style=apa,sorting=nyt,backend=biber,natbib]{biblatex}
\usepackage{lscape}
\usepackage{afterpage}
\usepackage{xcolor}
\usepackage{authblk}
\usepackage{multicol}
\newcommand{\pdiff}[2]{\frac{\partial#1}{\partial#2}}
\newcommand{\pdiffn}[3]{\frac{\partial^{#3}#1}{\partial#2^{#3}}}

\newcommand{\fdiff}[2]{\frac{\delta#1}{\delta#2}}

\newcommand{\dd}[1]{\mathrm{d}{#1}}
\newcommand{\ex}[1]{\mathrm{e}^{#1}}

\newcommand{\vw}[0]{\mathbf{w}}
\newcommand{\vA}[0]{\mathbf{A}}
\newcommand{\vr}[0]{\mathbf{r}}
\newcommand{\vc}[0]{\mathbf{c}}
\newcommand{\rhom}[0]{r_\mathrm{H}}
\newcommand{\chom}[0]{c_\mathrm{H}}
\newcommand{\mA}[0]{\mathbf{A}}
\newcommand{\mB}[0]{\mathbf{B}}
\newcommand{\mC}[0]{\mathbf{C}}
\newcommand{\mD}[0]{\mathbf{D}}
\newcommand{\mM}[0]{\mathbf{M}}
\newcommand{\mG}[0]{\mathbf{G}}
\newcommand{\real}[1]{\mathrm{Re}\left[#1\right]}
\newcommand{\imag}[1]{\mathrm{Im}\left[#1\right]}

\begin{document}

\title{Slow evolution towards generalism in a model of variable dietary range}
\author[1*]{Elliot M. Butterworth}
\author[1]{Tim Rogers}
\affil[1]{Department of Mathematical Sciences, University of Bath, Claverton Down, Bath, BA2 7AY, United Kingdom}
\affil[*]{Correspondence author.  Email: emb209@bath.ac.uk}
\date{}
\maketitle

\begin{abstract}
    Species sharing a habitat will co-evolve to make use of the available resources, as consumption is modulated by competition and negative feedback loops between consumers and resources. The dietary range of a given species determines the resources it has access to and thus the other species with which it competes. A narrow dietary range avoids competition at the cost of over-reliance on a small selection of resources; conversely a wide dietary range provides more alternatives but also more chance of competition with other species. Here, we investigate the evolution of dietary range within a mathematical model of niche formation. We find highly path dependent co-evolution dynamics characterised by long-lived quasi-stable states. Ultimately, stochastic effects drive the evolution of generalist diets, as we uncover in our analysis and simulations.
\end{abstract}

\section*{Keywords}
Pattern formation; Resource-consumer models; Evolution; Niche formation.

\section{Introduction}
Niche formation is a mechanism by which biodiversity can be maintained within an ecological system by circumventing competition between species for common resources \citep{Darwin1859,Grant2006,Letten2017}.  Species may adapt, for example, to make use of a distinct subset of resources little used by other species, so that they can obtain them with minimal conflict \citep{Dehling2021,Vamosi2014}.  Equally, species can adapt to use a broader range of resources, meaning they need not rely on any single one whose levels may fluctuate over time \citep{vonMeijenfeldt2023}.  Thus, ecologies can contain both generalists and specialists \citep[e.g.,][]{vonMeijenfeldt2023,Dehling2021}, although the extent to which each is preferred is likely dependent on many factors \citep{Dennis2011}, especially the heterogeneity of the environment \citep{Kassen2002}.  Although these resources can, in theory, be abiotic such as habitat niches, used for shelter or raising young, here we focus on resources in the sense of food that individuals must eat in order to survive.

\begin{figure}[htb!]
    \centering
    \includegraphics[width=\linewidth]{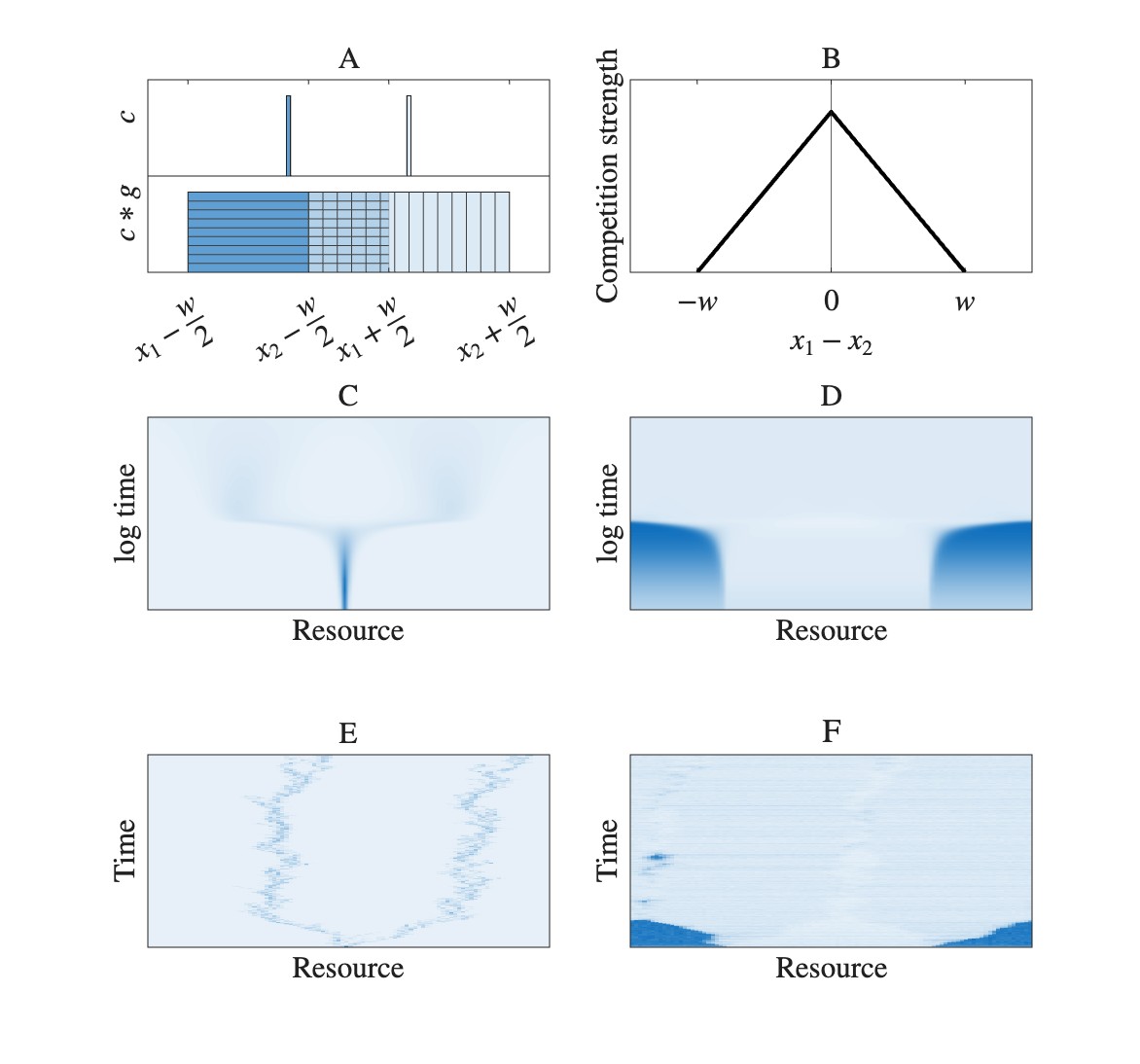}
    \caption{A: The competition between two species at positions \(x_1\) and \(x_2\) is given by the overlap in their resource preference distributions.  B: The competition kernel for this type of indirect competition is always positive definite, as it is the convolution of the resource preference distribution with itself.  Thus, species are not expected to form in the model.  C,D: This can be demonstrated with a numerical solution to the deterministic equations which slowly tends towards the homogeneous steady state, even from an initially heterogeneous distribution.  E,F: The demographic noise in stochastic simulations, however, causes patterns to persist, interpreted as species.  The pattern arising can be predicted from analytical analysis of the deterministic equations.  C and E show the consumer distributions and D and F show the resource distributions.}
    \label{fig:TheBigIssueFigure}
\end{figure}

When systems of competing individuals are reduced to simple mathematical models, predictable pattern formation has been shown to occur.  Work by \citet{MacArthur1967} on competition between species prompted research into Lotka-Volterra-type equations.  Assuming that species use resources according to some preference distribution, the overlap of two such distributions can be taken as the magnitude of the competition between two species, based on the assumption that this is related to the probability that those species encounter each other as they forage for resources \citep{Roughgarden1979,Levins1968}.  A simpler approach is to ignore such distributions of preference and assume instead that competition between two species can be quantified by a competition kernel, a function of the ``distance'' between each species' preferred resource.  The shape of this competition kernel is important as such models have been demonstrated to allow clustering, interpreted as the formation of species, only when the competition kernel is not positive-definite, meaning its Fourier transform has negative values \citep{Pigolotti2007,Pigolotti2010,HernandezGarcia2009,Leimar2013}.  When species formation does occur in simplified resource-consumer systems, it is often predictable and uniform.  This can be true even when the environment is heterogeneous \citep{HernandezGarcia2009,Barabas2009}, although probably for significantly simplified models only \citep{Szabo2006}.

The shape of a competition kernel can determine whether species are predicted to form in a given model.  In many systems, the type of competition is indirect, such as when consumers compete with each other over a common resource.  When this is the case, the resulting competition kernel is always positive definite \citep{Rogers2015}, as it is found by convolving the resource preference function (describing each species' preference for each resource) with itself.  This means that species are not predicted to evolve under this ubiquitous mode of competition.  One resolution to this is revealed when the effects of demographic noise are considered.  Such noise within a system can induce species formation even in cases where analytical predictions are to the contrary \citep{Rogers2012,Rogers2015}.

In this work, we investigate a continuous-space resource-consumer model, including the dynamics of the resources explicitly throughout our investigation.  Whilst resources are used to explain the competition between individuals in many resource-consumer models, rarely do they feature explicitly.  In our model, the resources would proliferate to a carrying capacity in the absence of consumers.  When present, each consumer depletes resources within its dietary range, defined as the total extent of its preference distribution, a function describing that consumer's preference for each resource.  The consumers have a natural birth and death rate, which are modified by consuming resources.  To allow the population of consumers to evolve, parameters of the dietary range can be modified between parent and offspring.  Initially, we allow the resources included in the dietary range to vary only, but subsequently extend the model to allow to vary the dietary range itself, permitting the evolution of specialised diets (narrow dietary range, strong preference for each resource in the range) and generalised diets (wide dietary range, low preference for each resource within the range).

For both models, we demonstrate that the only stable solution to the deterministic equations with a finite population is homogeneous, suggesting no species formation, but that demographic noise in stochastic simulations induces patterns which are interpreted as species formation.  In the case where dietary range is fixed, we demonstrate how the number of species can be predicted using Fourier analysis to reveal the dominant perturbations in the system linearised about its homogeneous steady state.  In the extended model with evolvable dietary range, we discover that dynamics occur on two timescales: species co-evolve rapidly to a state described by a quasi-stable manifold.  Analytical analysis of the deterministic system shows that it relaxes slowly along this manifold towards the homogeneous steady state.  As before, the presence of demographic fluctuations can prevent this relaxation, maintaining patterns in the system which lie within this manifold.  The system linearised about its homogeneous steady state can also be used to predict the most likely pattern formation in this extended model.  Demographic noise and especially the differential possibility of stochastic extinction for the species represented by various patterns mean certain states within the manifold are less stable than others.

\section{A model of consumers with fixed dietary range}
We begin with the scenario in which dietary range is not evolvable, but the resources included in the dietary range can evolve along a lineage.  This could be used to model a scenario such as the adaptive radiation of Darwin's Galapagos finches, where each species evolved the morphology to allow a distinct diet \citep{Lack1947,Bowman1961,Grant1999,Foster2008}.  The dynamics of the system are as follows: in the absence of consumers, the resources, whose distribution is denoted \(r(t,x)\), proliferate up to a carrying capacity \(\kappa\).  The growth rate is controlled by a parameter \(\alpha\).  The resources are depleted by the consumers: the amount by which each type of resource is depleted is dependent on the consumer population and the preference distribution, \(g(x)\).  The consumer population, whose distribution is denoted \(c(t,x)\), has a natural birth and death rate (\(\beta\) and \(\delta\) respectively).  Having available resources modifies the birth rate, with effect size controlled by the parameter \(\gamma\).  Evolution along lineages occurs via consumer ``mutation'': offspring and parent differ in their position along the resource axis, \(x\), by a distance whose probability distribution is Gaussian.  This mutation rate is controlled by the parameter \(\mu\) (related to the variance of the Gaussian distribution, see Appendix~\ref{sec:DerivationOfPDEs}).

In Appendix~\ref{sec:DerivationOfPDEs} we use this microscopic description to derive a pair of partial differential equations (Equations \eqref{eq:ResourcePDE} and \eqref{eq:ConsumerPDE}) in the limit of a large population such that demographic noise is negligible and the system becomes deterministic.  They are
\begin{align}\label{eq:ResourcePDE}
    \frac{\partial r}{\partial t}&=\alpha\left(1-\frac{r}{\kappa}\right)-r(c*g),\\\label{eq:ConsumerPDE}
    \frac{\partial c}{\partial t}&=(\beta-\delta)c+\gamma c(r*g)+\mu\nabla^2c.
\end{align}
The method of the derivation is to first write down a master equation involving operators which change the state of the system, defined by the distribution of resources and consumers at a given time.  We show how the operators can be expanded in a series of increasing order of the reciprocal of a large parameter related to the number of resources or consumers.  Taking only the zeroth order terms, equivalent to allowing the system size to become infinite such that all demographic effects are lost, yields deterministic partial differential equations.

To begin the analysis of the deterministic equations, we look for homogeneous steady states of the system.  Since there is diffusion in Equation~\eqref{eq:ConsumerPDE}, the only steady state of the system will be homogeneous in \(x\), implying no species form in this model.  We will denote the homogeneous steady states for the resources and consumers \(\rhom\) and \(\chom\) respectively.  Throughout, we will work with periodic boundaries on \(x\in[0,L)\) such that edge effects do not have to be considered.  In Appendix~\ref{sec:NeumannBoundaries} we demonstrate that the dynamics are not greatly affected by making the resource domain non-periodic.  One consequence of the periodic boundary conditions is that the convolution of \(\rhom\) or \(\chom\) with the resource preference distribution returns the same constant.  Noting this and setting the time derivative equal to zero, we produce the equations defining the homogeneous steady states:
\begin{equation*}
    \alpha\left(1-\frac{\rhom}{\kappa}\right)-\rhom\chom=0,\qquad(\beta-\delta)\chom+\gamma\chom\rhom=0.
\end{equation*}
The non-trivial solution is
\begin{equation*}
    \rhom=\frac{\delta-\beta}{\gamma},\qquad\chom=\alpha\left(\frac{\gamma}{\delta-\beta}-\frac{1}{\kappa}\right).
\end{equation*}
Note that only when \(\delta>\beta\) do we obtain a positive value of \(\rhom\).  Thus, for the parameter regime we will be interested in, the consumer population shrinks in the absence of resources to consume.  Note also that another steady homogeneous solution to the equations is \(\rhom=\kappa\), \(\chom=0\).  We are not interested in this extinction state of the consumers.

We can look at the linearised dynamics of the fluctuations about this homogeneous steady state.  Following the method in \cite{ButlerGoldenfeld2009, Butler2011} we let \(r=\rhom+\tilde{r}\) and \(c=\chom+\tilde{c}\), where \(\tilde{r}\) and \(\tilde{c}\) are small perturbations to the homogeneous states.  Substituting these expressions into Equations \eqref{eq:ResourcePDE} and \eqref{eq:ConsumerPDE} gives
\begin{align*}
    \pdiff{\rhom+\tilde{r}}{t}&=\alpha\left(1-\frac{\rhom+\tilde{r}}{\kappa}\right)-(\rhom+\tilde{r})[(\chom+\tilde{c})*g],\\
    \pdiff{\chom+\tilde{c}}{t}&=(\beta-\delta)(\chom+\tilde{c})+\gamma(\chom+\tilde{c})[(\rhom+\tilde{r})*g]+\mu\nabla^2(\chom+\tilde{c}).
\end{align*}
Using the definitions of the homogenous steady state, we can remove the terms involving no small quantities, yielding
\begin{align*}
    \pdiff{\tilde{r}}{t}&=-\frac{\alpha\tilde{r}}{\kappa}-\tilde{r}\chom-\rhom(\tilde{c}*g)+\mathcal{O}(\tilde{r}\tilde{c}),\\
    \pdiff{\tilde{c}}{t}&=(\beta-\delta)\tilde{c}+\gamma\tilde{c}\rhom+\gamma\chom\tilde{r}*g+\mu\nabla^2\tilde{c}+\mathcal{O}(\tilde{c}\tilde{r}).
\end{align*}
Ignoring the terms which are second order in small quantities and taking the Fourier transform of the equations in both space and time, we find
\begin{align*}
    -i\omega\tilde{r}_k&=-\frac{\alpha\tilde{r}_k}{\kappa}-\tilde{r}_k\chom-\rhom\tilde{c}_k\mathcal{G}_k,\\
    -i\omega\tilde{c}_k&=(\beta-\delta)\tilde{c}_k+\gamma\tilde{c}_k\rhom+\gamma\chom\tilde{r}_k\mathcal{G}_k-\mu k^2\tilde{c}_k.
\end{align*}
The Fourier transform of \(\tilde{r}\) and \(\tilde{c}\) is \(\tilde{r}_k\) and \(\tilde{c}_k\) respectively and \(\mathcal{G}_k\) is the Fourier transform of the preference distribution \(g(x)\).  Using the definitions of \(\rhom\) and \(\chom\), these equations simplify to
\begin{equation*}
    -i\omega\tilde{r}_k=-\frac{\alpha}{\rhom}\tilde{r}_k-\rhom\mathcal{G}_k\tilde{c}_k,\qquad-i\omega\tilde{c}_k=\gamma\chom\mathcal{G}_k\tilde{r}_k-\mu k^2\tilde{c}_k.
\end{equation*}
Defining \(\boldsymbol{\phi}=(\tilde{r}_k,\tilde{c}_k)^T\), we can rewrite these two equations as a single matrix equation
\begin{equation*}
    -i\omega\boldsymbol{\phi}=\begin{bmatrix}
        -\frac{\alpha}{\rhom}&-\rhom\mathcal{G}_k\\
        \gamma\chom\mathcal{G}_k&-\mu k^2
    \end{bmatrix}\boldsymbol{\phi}\equiv\mathbf{A}\boldsymbol{\phi}.
\end{equation*}
Linear stability of the homogeneous state can be assessed by examining the eigenvalues of the matrix \(\mA\).  The eigenvalue with the the maximum real part across all values of \(k\) corresponds to the longest lived perturbed mode.  Those eigenvalues are
\begin{equation*}
    \lambda_\pm=\frac{1}{2}\left[-\left(\frac{\alpha}{\rhom}+\mu k^2\right)\pm\sqrt{\left(\frac{\alpha}{\rhom}-\mu k^2\right)^2-4\gamma\rhom\chom\mathcal{G}_k^2}\right].
\end{equation*}
Since \(\frac{\alpha}{\rhom}+\mu k^2\geq\left|\frac{\alpha}{\rhom}-\mu k^2\right|\) and \(4\gamma\rhom\chom\mathcal{G}_k^2\geq0\) (for real \(\mathcal{G}_k\)), we see that \(\real{\lambda_\pm}\leq0\) and so any perturbation from the homogeneous steady state decays over time.  Thus, in the deterministic setting, the homogeneous steady state (no species predicted) is stable.  This can be seen in the middle panels in Figure~\ref{fig:TheBigIssueFigure} where, despite an initially inhomogeneous distribution of consumers, the perturbation dies away over time as the system tends towards a homogeneous state.

\begin{figure}[htb!]
    \centering
    \includegraphics[width=\textwidth]{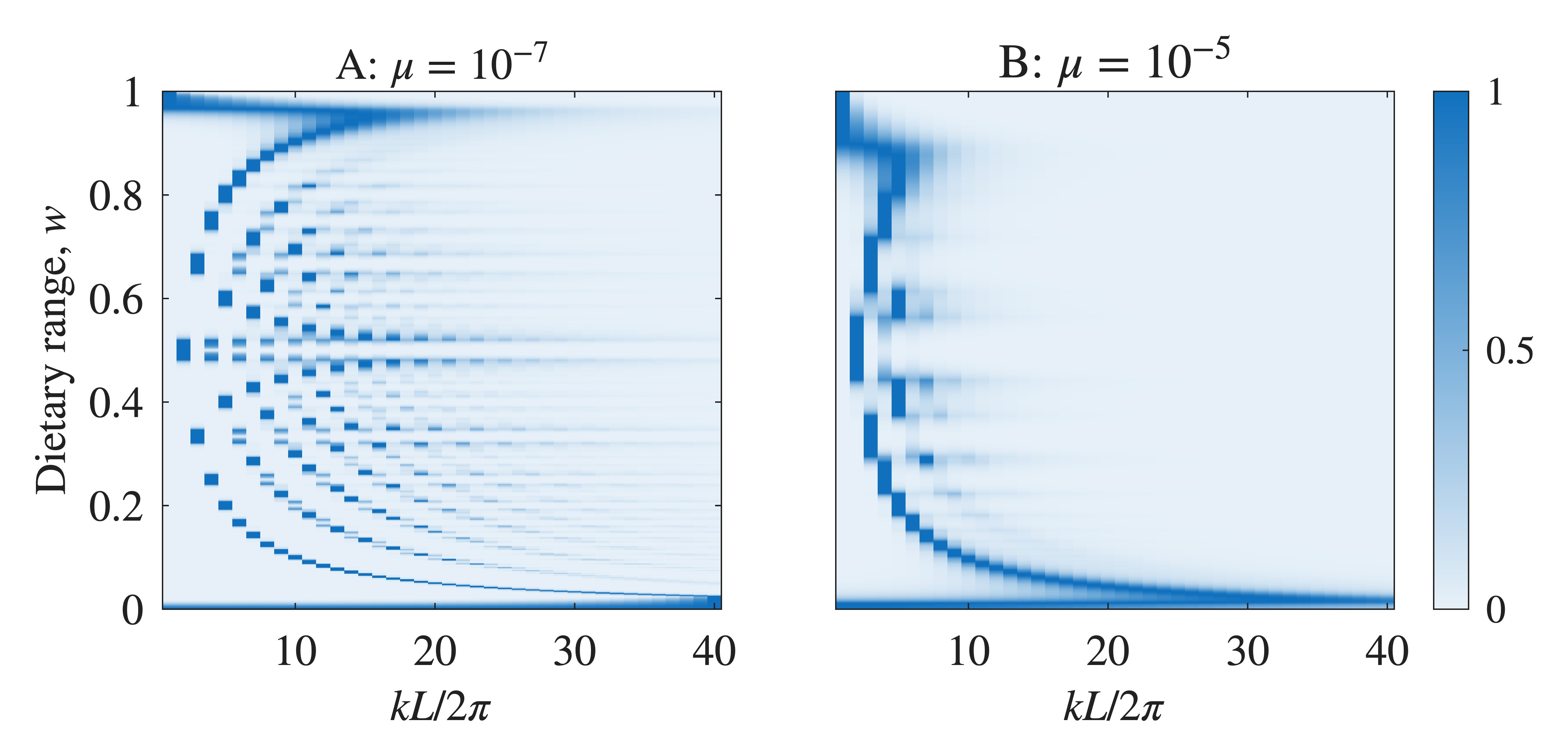}
    \caption{The power spectrum \(\left<|\tilde{c}_k|^2\right>\) for various (integer) values of \(\frac{kL}{2\pi}\) between zero and 40 and values for the dietary range, \(w\), between zero and one.  For each value of \(w\), the power spectrum was normalised by dividing all values of the spectrum by the power of the highest peak.  The parameter values for the system were \(\alpha=2\times10^4\), \(\beta=1\), \(\delta=2\), \(\gamma=0.01\), \(\kappa=1\times10^5\) and \(L=1\).  In A, \(\mu=10^{-7}\) and in B, \(\mu=10^{-5}\).}
    \label{fig:PowerSpectrum_0<w<1}
\end{figure}

Although the deterministic system always tends towards a homogenous state, stochastic simulations of the same system, but with a finite population, produce highly inhomogeneous distributions of consumers that persist over time, as seen in the bottom panels of Figure~\ref{fig:TheBigIssueFigure}.  To understand this, we wish to investigate the effects of demographic noise, randomness introduced due to a finite populations and the associated random timing of all events which occur within the population.  As an approximation to this intrinsic noise and in line with other studies \citep[see;][and references therein]{Butler2011}, we now add external noise, \(\boldsymbol{\xi}=(\xi_1,\xi_2)^T\), to the system:
\begin{equation*}
    (\mathbf{A}+i\omega)\boldsymbol{\phi}=-\boldsymbol{\xi}.
\end{equation*}
Thus,
\begin{equation*}
    \boldsymbol{\phi}=-|\mathbf{A}+i\omega|^{-1}\begin{bmatrix}
        -\mu k^2+i\omega&\rhom\mathcal{G}_k\\
        -\gamma\chom\mathcal{G}_k&-\frac{\alpha}{\rhom}+i\omega
    \end{bmatrix}\boldsymbol{\xi}.
\end{equation*}
We can find the fluctuations in the consumer numbers around the homogeneous steady state.  Assuming uncorrelated noise (\(\left<\xi_1\xi_2\right>=0\)), since no event changes the resource and consumer numbers simultaneously, we obtain the power spectrum
\begin{equation}\label{eq:PowerSpectrumConstantGrowth}
    \left<|\tilde{c}_k|^2\right>=|\mathbf{A}+i\omega|^{-2}\left((\gamma\chom\mathcal{G}_k)^2\left<\xi_1\xi_1\right>+\left(\frac{\alpha}{\rhom}\right)^2\left<\xi_2\xi_2\right>\right).
\end{equation}
The square of the determinant is
\begin{multline*}
    |\mathbf{A}+i\omega|^2=\left(\frac{\alpha}{\rhom}\mu k^2-\omega^2\right)^2+\omega^2\left(\frac{\alpha}{\rhom}+\mu k^2\right)^2+\left(\gamma\rhom\chom\right)^2\left|\mathcal{G}_k^2\right|^2+\\2\gamma\rhom\chom\left(\left(\frac{\alpha}{\rhom}\mu k^2-\omega^2\right)\real{\mathcal{G}_k^2}-\omega\left(\frac{\alpha}{\rhom}+\mu k^2\right)\imag{\mathcal{G}_k^2}\right).
\end{multline*}
Peaks in the power spectrum indicate perturbed modes which dominate the system and so allow us to predict patterns which will emerge.  Figure~\ref{fig:PowerSpectrum_0<w<1} shows how this dominant mode changes as the dietary range, \(w\), is varied between 0 and 1.  Note that a qualitatively identical figure can be produced by looking at maxima in a plot of \(\lambda_\mathrm{max}=\mathrm{max}(\lambda_+,\lambda_-)\) against \(k\) that shows which perturbations are most long lived.  These long lived modes can be used to predict the number of distinct clusters forming in a stochastic population.

\begin{figure}[htb!]
    \centering
    \includegraphics[width=\textwidth]{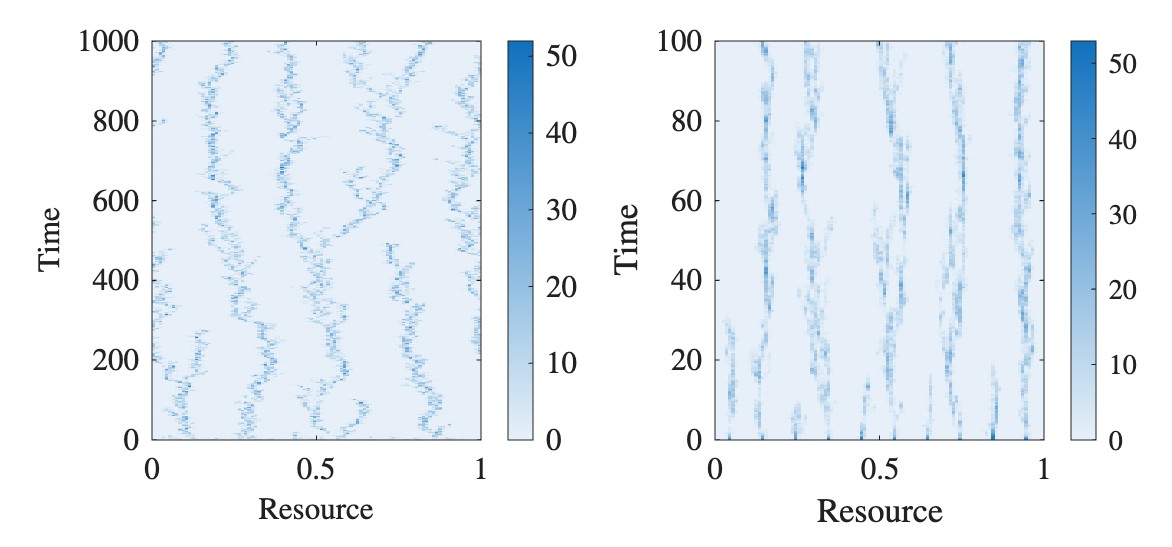}
    \caption{The evolution of a system containing individuals that can evolve their preferred resources (moving along the resource axis \(x\) (100 bins)) and whose dietary range is \(w=0.2\).  A: Initially there are five consumers in each bin along \(x\).  Five approximately equally spaced species form rapidly and despite noise causing various speciation and extinction events, the system recovers to a state with five species equally spaced along \(x\).  B: Initially there were 10 equally spaced species in this system.  Once an extinction event happens (which is very soon in the simulation) the system is unlikely to recover to the initial state, instead evolving rapidly to the state with five equally spaced species.  The parameters of this system were \(\alpha=2\times10^4\), \(\beta=1\), \(\delta=2\), \(\gamma=0.01\), \(\kappa=1\times10^5\), \(\mu=10^{-5}\) and \(L=1\).}
    \label{fig:RCSimulationFromHomo_w02}
\end{figure}

The propensity for this resource-consumer system to diverge from its homogeneous steady state can be demonstrated by running stochastic simulations of the dynamics.  An example is shown in Figure~\ref{fig:RCSimulationFromHomo_w02}, A.  The preference distribution in this simulation was a bounded uniform function,
\begin{equation*}
    g(x)=\begin{cases}
        \frac{1}{w}&\text{if }-\frac{w}{2}\leq x\leq\frac{w}{2},\\
        0&\text{otherwise},
    \end{cases}
\end{equation*}
with dietary range \(w=0.2\) (20\% of all the resources were available to each consumer).  We see that five, approximately equally spaced species form in the population.  This is the pattern predicted by the power spectrum, which has its highest peak at five for \(w=0.2\) (Figure~\ref{fig:PowerSpectrum_0<w<1}).  This pattern makes intuitive sense: five equally spaced species, each of which has access to 20\% of all resources, make use of the full set of resources to an equal extent (note that the population size of each species is approximately equal) without any competition for those resources from other species.  The same would be true of 10 regularly spaced species, except that in this case, there would be equal competition between two species for each resource.  Note also that since there are twice as many species in this pattern, but the rate of resource production is the same, the population size of each species supported by the resources is halved, meaning that each species is more at risk of stochastic extinction.  Thus, this pattern of resource allocation is short lived (Figure~\ref{fig:RCSimulationFromHomo_w02}, B) and does not appear spontaneously in stochastic simulations.

\section{A model of consumers with evolvable dietary range}
We now extend the model presented above to allow dietary range, \(w\), to evolve along a lineage.  This allows for the evolution of population members specialised for obtaining specific foods within a narrow range of resource space (\(w\ll L\)) as well as the evolution of more generalist strategies, meaning consumers make use of a wide variety of resources (\(w\approx L\)).  We will allow dietary range to take values in the range \(w\in[w_1,w_2]\), permitting the evolution of individuals with any value in this range.  The most extreme example (for a bounded uniform preference distribution) would be \(w\in[0,L]\), that is, the preference distribution can vary between a delta function, suggesting only a single resource can be accessed by a consumer with an infinitesimally narrow dietary range (a monophagous consumer), and a uniform distribution, suggesting the consumer has no preference for any resource over another.  This consideration modifies Equations \eqref{eq:ResourcePDE} and \eqref{eq:ConsumerPDE} as follows:
\begin{align}\label{eq:ResourcePDEModified}
    \pdiff{r}{t}&=\alpha\left(1-\frac{r}{\kappa}\right)-r\left(c*g\right),\\\label{eq:ConsumerPDEModified}
    \pdiff{c}{t}&=\left(\beta-\delta\right)c+\gamma c\left(r*g\right)+\nabla_\mu^2c.
\end{align}
We have defined the operator
\begin{equation*}
    \nabla_\mu^2\equiv\mu_x\pdiffn{}{x}{2}+\mu_w\pdiffn{}{w}{2}.
\end{equation*}
As before, \(x\) represents resource space and \(w\) represents dietary range, such that \(r=r(t,x)\), \(c=c(t,x,w)\) and \(g=g(x,w)\).  Necessarily, the convolution terms are modified to
\begin{align*}
    (c*g)(x)&=\int_{w_1}^{w_2}\int_{-\infty}^\infty c(t,z,w)g(x-z,w)\dd{z}\dd{w},\\
    (r*g)(x,w)&=\int_{-\infty}^\infty r(t,z)g(x-z,w)\dd{z}.
\end{align*}
Note that they are now asymmetric from the point of view of the resources and consumers to take into account the different dimensional spaces the two distributions are defined over.  We have also introduced a parameter to allow the mutation rate in the \(x\) and \(w\) directions to differ.  We are able to investigate scenarios where either resource preference evolves far more rapidly than dietary range, \textit{vice versa}, and intermediate scenarios.

We find the homogeneous steady states as before, setting the derivatives of the resource and consumer distributions to zero.  The resulting homogeneous steady state values are
\begin{equation*}
    \chom=\begin{cases}
        \frac{\alpha}{w_2-w_1}\left(\frac{\gamma}{\delta-\beta}-\frac{1}{\kappa}\right)&\text{if }\rhom=\frac{\delta-\beta}{\gamma},\\
        0&\text{if }\rhom=\kappa.
    \end{cases}
\end{equation*}
We will, as before, ignore the extinction state of the consumer population.  Note that in the finite population state, the term \((w_2-w_1)^{-1}\) becomes singular as \((w_2-w_1)\rightarrow0\), demonstrating that \(c\) should be interpreted as a density over dietary range values, meaning consumer numbers are found by integrating \(c(t,x,w)\) over \(w\).  Note also that in the case when the dietary range is not allowed to evolve, we have fixed dietary range \(w=w_1=w_2\), such that this scenario is recovered in the limit \(w_1\rightarrow w_2\).

We now consider small perturbations around the non-trivial homogeneous steady state: \(r(t,x)=\rhom+\tilde{r}(t,x)\) and \(c(t,x,w)=\chom+\tilde{c}(t,x,w)\).  Substituting these expressions into Equations \eqref{eq:ResourcePDEModified} and \eqref{eq:ConsumerPDEModified} and removing terms which sum to zero by the definition of the homogeneous steady state, we find the equations
\begin{align*}
    &\pdiff{\tilde{r}}{t}=-\frac{\alpha}{\rhom}\tilde{r}-\rhom(\tilde{c}*g)+\mathcal{O}(\tilde{r}\tilde{c}),\\
    &\pdiff{\tilde{c}}{t}=(\beta-\delta)\tilde{c}+\gamma\rhom\tilde{c}+\gamma\chom(\tilde{r}*g)+\nabla_\mu^2\tilde{c}+\mathcal{O}(\tilde{r}\tilde{c}).
\end{align*}
To obtain linear equations describing the time evolution of small perturbations to the homogeneous steady state, we will ignore terms which are second order in small quantities.  We now consider the Fourier transform of \(\tilde{r}\) and \(\tilde{c}\) in resource space, \(x\) only, in contrast to our previous approach where we applied a Fourier transform in time also.  The transformed functions are
\begin{align*}
    \tilde{R}_k&=\tilde{R}_k(t)=\mathcal{F}_k\left[\tilde{r}\right]=\int_{-\infty}^{\infty}\ex{-ikx}\tilde{r}(t,x)\dd{x},\\
    \tilde{C}_k&=\tilde{C}_k(t,w)=\mathcal{F}_k\left[\tilde{c}\right]=\int_{-\infty}^{\infty}\ex{-ikx}\tilde{c}(t,x,w)\dd{x},\\
    G_k&=G_k(w)=\mathcal{F}_k[g]=\int_{-\infty}^\infty\ex{-ikx}g(x,w)\dd{x}.
\end{align*}
By the convolution theorem, the convolution of each perturbation with the preference distribution becomes the product of their Fourier transforms.  The second derivative with respect to \(x\) is also replaced by a term involving \(k^2\).  Using these results and definitions lead to the following equations for \(\tilde{R}_k(t)\) and \(\tilde{C}_k(t)\):
\begin{equation}\label{eq:tEvolutionRTilde_k}
    \pdiff{\tilde{R}_k(t)}{t}=-\frac{\alpha}{\rhom}\tilde{R}_k(t)-\rhom\int_{w_1}^{w_2}\tilde{C}_k(t,w)G_k(w)\dd{w}
\end{equation}
and
\begin{multline}\label{eq:tEvolutionCTilde_k}
    \pdiff{\tilde{C}_k(t,w)}{t}=\left(\beta-\delta+\gamma\rhom\right)\tilde{C}_k(t,w)+\gamma\chom\tilde{R}_k(t)G_k(w)-\\
    \mu_xk^2\tilde{C}_k(t,w)+\mu_w\pdiffn{\tilde{C}_k(t,w)}{w}{2}.
\end{multline}

Although preferable for analytical investigation, implementing periodic boundary conditions on \(w\) would allow highly specialised individuals (\(w\approx0\)) to produce completely generalist offspring (\(w\approx1\)) and completely generalist individuals to produce highly specialised offspring.  To avoid this biologically unrealistic event, we implement Neumann (reflecting) boundaries at \(w_1\) and \(w_2\) so mass is not lost from the system.  However, this does mean that moving into the Fourier domain in dietary range space is not useful and we require another method to remove the integral in Equation~\eqref{eq:tEvolutionRTilde_k}, allowing us to find the dominant perturbations about the homogeneous steady state.  To this end, we coarse-grain \(w\) to allow this integral to be rewritten as a matrix product.  The steps of this analysis are shown in Appendix~\ref{sec:CoarseGrainingDietaryRange}.  The result is a matrix whose eigenvalues describe which perturbed modes die away most slowly and so indicate, via their corresponding eigenvectors, the linear prediction for the number of species we expect to see forming in stochastic simulations.

\begin{figure}[htb!]
    \centering
    \includegraphics[width=\textwidth]{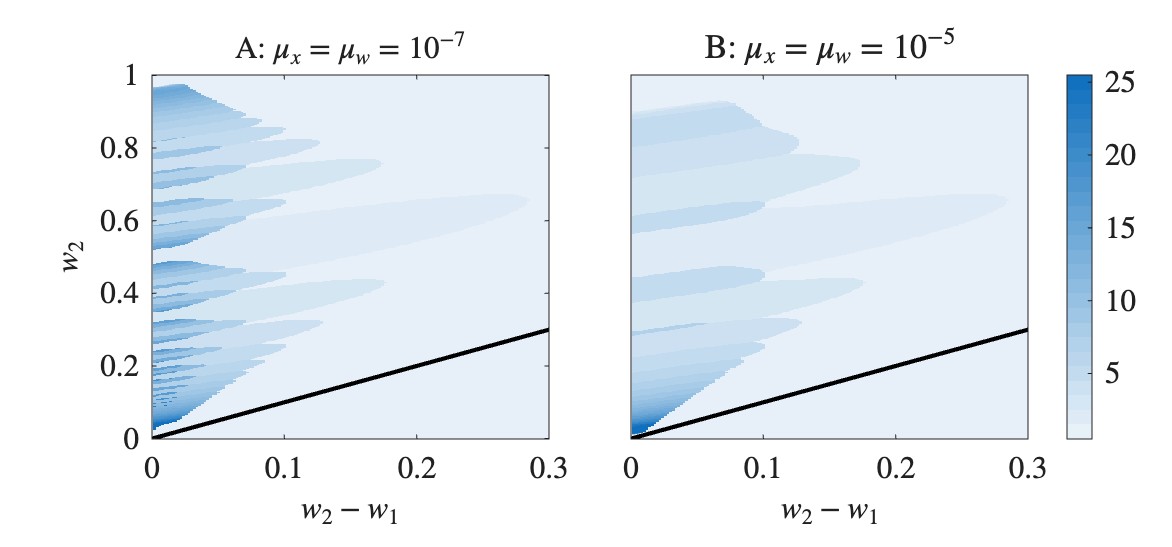}
    \caption{The dominant pattern-forming mode for different combinations of \(w_2\) and \(w_2-w_1\).  In the case where \(w_1\rightarrow w_2\) the results from the analysis without evolvable dietary range are recovered.  As the width of the window of permitted dietary ranges expands, the higher-\(k\) modes become less dominant until the first mode dominates (interpreted as a single species).  The parameter values for this system were \(\alpha=2\times10^4\), \(\beta=1\), \(\delta=2\), \(\gamma=0.01\), \(\kappa=1\times10^5\) and \(L=1\).  In A \(\mu_x=\mu_w=10^{-7}\).  In B \(\mu_x=\mu_w=10^{-5}\).  We see that greater noise leads to loss of some of the finer structure of this plot and the higher modes (i.e., the coexistence of a greater number of species).  The vector \(\vw\) contained 100 entries, giving a \(101\times101\) matrix to invert in each case (see Appendix~\ref{sec:CoarseGrainingDietaryRange} for details).  Note that the area below the black line \(w_2=w_2-w_1\) (i.e., \(w_1=0\)) is not meaningful for this system.  The entire area to the right of the region shown has dominant mode equal to one.}
    \label{fig:DominantEigenmodeFory2VSy1}
\end{figure}

\begin{figure}[htb!]
    \centering
    \includegraphics[width=\textwidth]{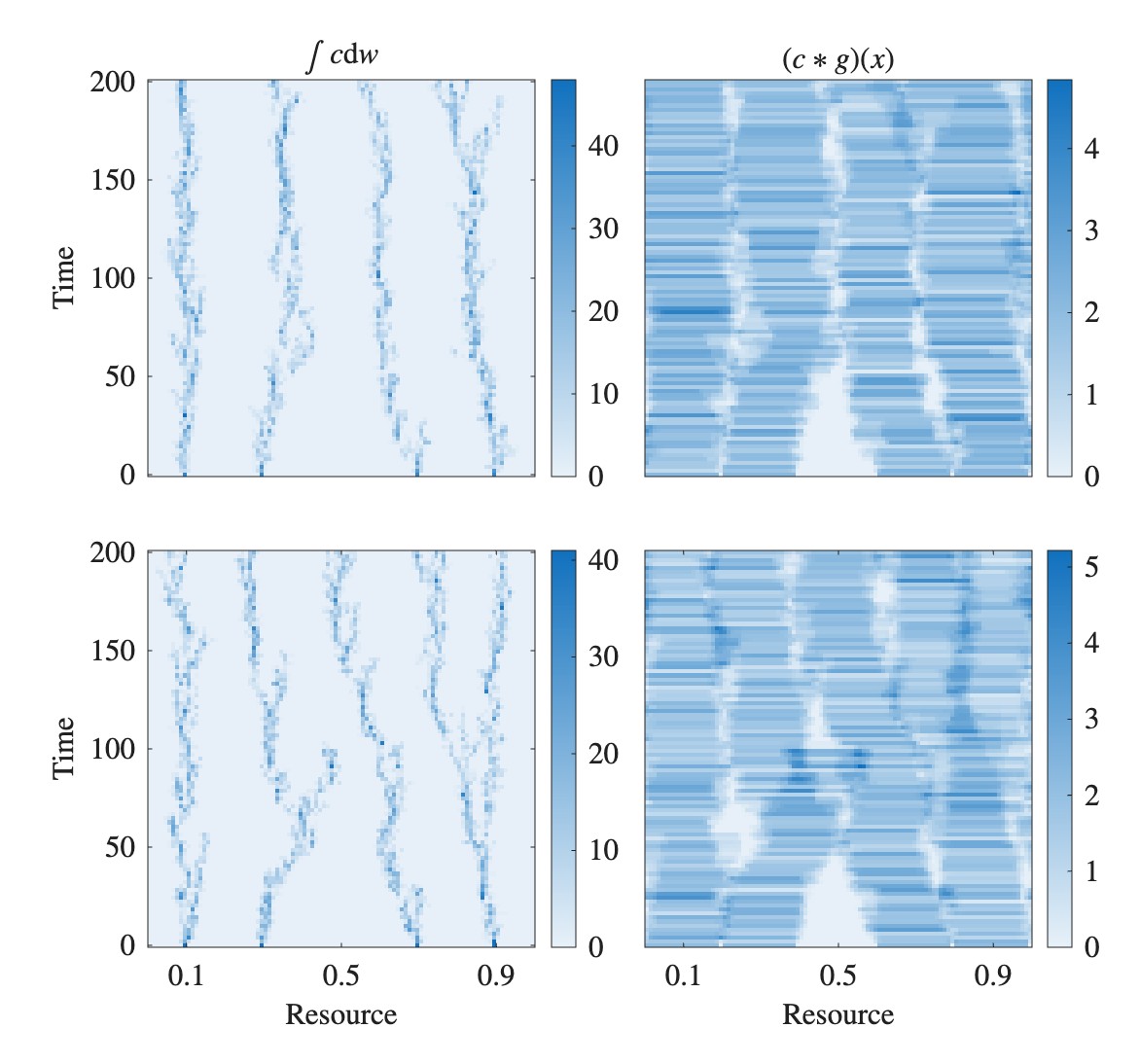}
    \caption{The system in both rows is in the same state at \(t=0\): five species with dietary range \(w=0.2\), equally spaced along the resource axis, \(x\) (100 bins), at the moment the centrally positioned species goes extinct.  The upper panels show the evolution of this system when \(0<w<1\) (50 bins), and the lower panels show the evolution when \(0.19<w<0.21\) (5 bins): an unrestricted and a restricted evolution of dietary range scenario.  The upper panels show an increase in dietary range on average to satisfy Equation~\eqref{eq:FlatConvolutionCondition} with four heterogeneous species (note the different widths of the ``columns'' in the right hand plot).  The lower panels show that the only way to satisfy the condition given the restrictions on \(w\) is with five species, equally spaced in \(x\).  The parameters of these systems were \(\alpha=2\times10^4\), \(\beta=1\), \(\delta=2\), \(\gamma=0.01\), \(\kappa=1\times10^5\), \(\mu_x=\mu_w=10^{-5}\) and \(L=1\).}
    \label{fig:FullVsRestricted_y_FollowingExtinction}
\end{figure}

Figure~\ref{fig:DominantEigenmodeFory2VSy1} shows the slowest decaying mode for different values of \(w_1\) and \(w_2\).  This Figure~was generated using the numerical eigenvalues of the matrix related to the time evolution of \(r\) and \(c\) (see Appendix~\ref{sec:CoarseGrainingDietaryRange} for details).  The highest value mode permitted was \(n=\frac{kL}{2\pi}=25\).  The Figure~shows patches of higher modes dominating when the \(w_2-w_1\) is restricted significantly enough.  In the limit of \(w_2-w_1\rightarrow0\), the results from the analysis with fixed dietary range are recovered.  The (approximate) line of symmetry down the centre of each patch has a positive gradient because the same mode (\(n\)) dominates when  \(\exists w^*\in[w_1,w_2]\) such that \(w^*=\frac{z}{n}\) for \(z\in\mathbb{N}\). That is, an integer number of dietary ranges fully span resource space an integer number of times.  For regularly spaced species of equal population sizes along the resource axis, this means that each resource is consumed to the same extent and the competition for each resource is equal.

\begin{figure}[htb!]
    \centering
    \includegraphics[width=\textwidth]{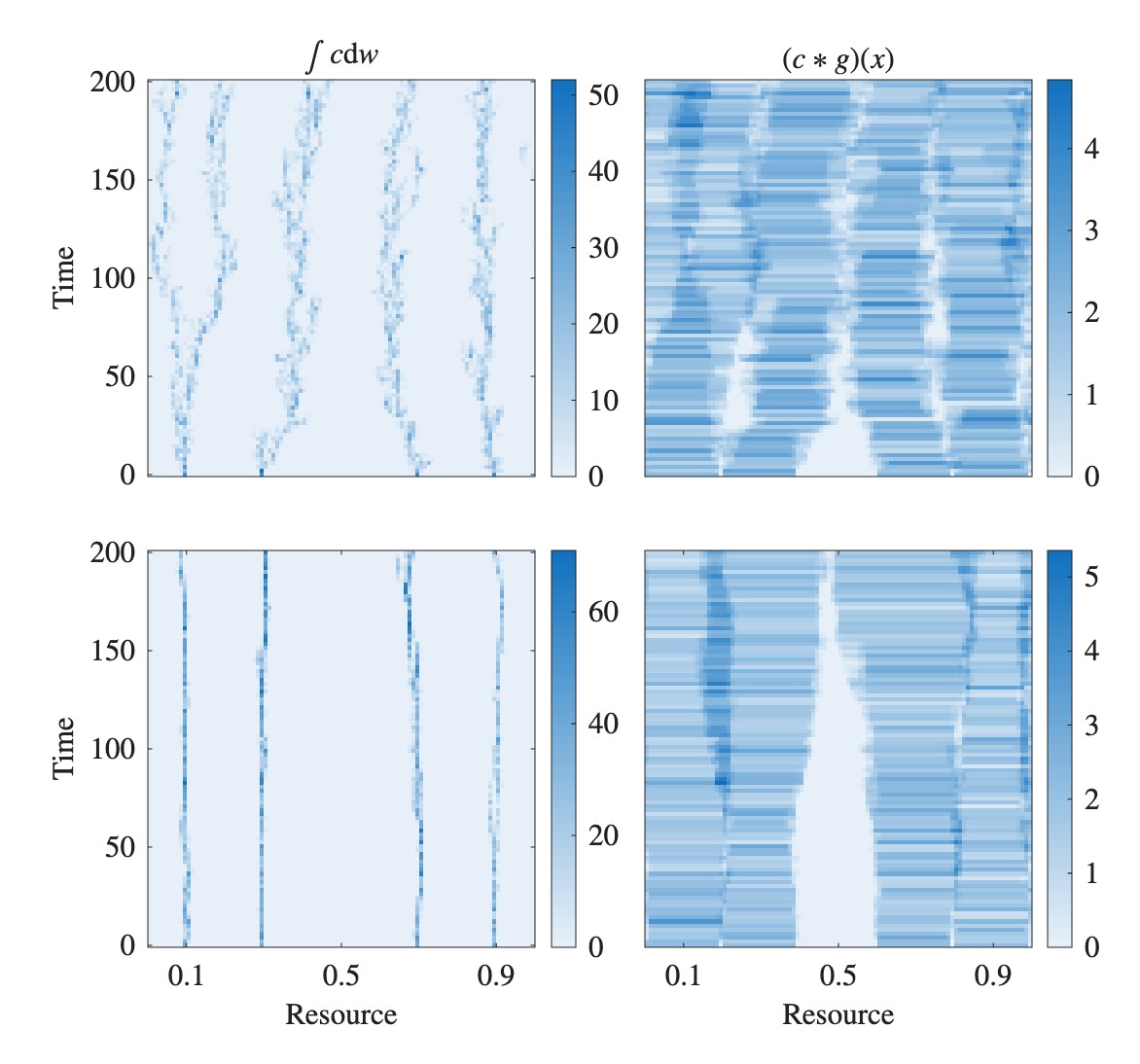}
    \caption{Two simulations with identical parameters and initial condition to the one in the upper panels of Figure~\ref{fig:FullVsRestricted_y_FollowingExtinction} except for the values of \(\mu_x\) and \(\mu_w\).  In the upper panels, \(\mu_x=10^{-5}\) and \(\mu_w=10^{-6}\), whereas in the lower panel, \(\mu_x=10^{-6}\) and \(\mu_w=10^{-5}\).  Thus, we see the difference in the evolution of a population when either position in resource space or dietary range is far more rapidly evolved.  In the former case, another species is produced to satisfy Equation~\eqref{eq:FlatConvolutionCondition}, but in the latter case it is satisfied by adjustments to the dietary range of each surviving species.}
    \label{fig:EvolvePositionVSEvolveWidthFollowingExtinction}
\end{figure}

Although Figure~\ref{fig:DominantEigenmodeFory2VSy1} only predicts patterns of more than one species for small values of \(w_2-w_1\), the results highlight a more general point.  If all resources are consumed at the same ``sustainable'' rate (such that they remain at their homogeneous steady level), it means that the distribution produced by convolving the population of consumers with their preference distributions is uniform and equivalent to the distribution produced when the consumer population is also at its homogeneous steady level.  Further, when all resources are consumed at this sustainable rate, the terms
\begin{equation*}
    \left(\beta-\delta\right)c(t,x,w)+\gamma c(t,x,w)\rhom=0,
\end{equation*}
meaning that the time evolution of the consumer population is given by
\begin{equation*}
    \pdiff{c}{t}=\nabla_\mu^2c.
\end{equation*}
Thus, assuming \(\mu_x,\mu_w\ll\beta,\delta,\gamma\rhom\), the time evolution is much slower in this regime (a quasi-steady regime).  Another consequence of \(r(t,x)=\rhom\) is
\begin{equation*}
    \pdiff{\rhom}{t}=0=\alpha\left(1-\frac{\rhom}{\kappa}\right)-\rhom(c*g)(t,x),
\end{equation*}
which rearranges to give
\begin{equation}\label{eq:FlatConvolutionCondition}
    (c*g)(t,x)=\alpha\left(\frac{1}{\rhom}-\frac{1}{\kappa}\right)=\chom(w_2-w_1).
\end{equation}
That is, the quasi-steady regime is realised for any consumer population which is the preimage of homogeneous steady state.  This defines a manifold of consumer populations which will evolve slowly due to (small) diffusive effects only.  All of the regular patterns implied by Figure~\ref{fig:DominantEigenmodeFory2VSy1} are preimages of the homogeneous steady state.  This manifold, however, also contains an infinite number of non-regular consumer population structures.  Such populations can be formed of both generalist and specialist species (here, defined by large and small \(w\) values respectively).  Consider, for example, a collection of point species:
\begin{equation*}
    c(x,w)=\sum_iA_i\delta(x-x_i)\delta(w-w^i),
\end{equation*}
for which
\begin{equation*}
    (c*g)(x)=\sum_i\begin{cases}
        \frac{A_i}{w^i}&\text{if }x_i-\frac{w^i}{2}\leq x\leq x_i+\frac{w^i}{2},\\
        0&\text{otherwise}.
    \end{cases}
\end{equation*}
\(A_i\) is the number of individuals of species \(i\) and \(w^i\) is the dietary range of that species (not to be confused with \(w_1\) and \(w_2\), the minimum and maximum values of dietary range which are evolutionarily accessible).  This population can exist within the quasi-steady manifold in an infinite number of arrangements which satisfy Equation~\eqref{eq:FlatConvolutionCondition}.  For example, if there is no competition (i.e., no overlap in resource preference), we require
\begin{equation*}
    \sum_iw^i=L
\end{equation*}
and spacing so that
\begin{equation*}
    x_{i+1}-x_i=\frac{1}{2}(w^{i+1}+w^i)
\end{equation*}
where the ordering of points is arbitrary (and the periodic boundary is respected).  Figures \ref{fig:FullVsRestricted_y_FollowingExtinction} and \ref{fig:EvolvePositionVSEvolveWidthFollowingExtinction} shows examples of consumer populations made from distinct species in \((x,w)\) (clusters of individuals occupying a narrow range of bins only) that evolve to become preimages of the homogeneous steady state, satisfying Equation~\eqref{eq:FlatConvolutionCondition}.  Following the extinction, there is a range of unused resources, identified by range of \(x\) values over which \((c*g)(x)=0\).  In each simulation shown in Figures \ref{fig:FullVsRestricted_y_FollowingExtinction} and \ref{fig:EvolvePositionVSEvolveWidthFollowingExtinction}, we see the consumer population evolve such that no resources are left unused.  Equally the species coevolve to maintain little to no competition for any resources, as shown by the minimal overlap of the ``columns'' in the right hand plots of Figures \ref{fig:FullVsRestricted_y_FollowingExtinction} and \ref{fig:EvolvePositionVSEvolveWidthFollowingExtinction}.  Thus, the species coevolve to a point where Equation~\eqref{eq:FlatConvolutionCondition} is satisfied, up to demographic fluctuations in the numbers of each species.  Although the exact arrangement of species and their diets is highly path dependent, the observation of rapid evolution to the quasi-steady manifold and subsequent slow variation within it, driven by demographic fluctuations, is predictable.

Another requirement for a point population with no overlapping preference distributions to satisfy Equation~\eqref{eq:FlatConvolutionCondition} is
\begin{equation}\label{eq:SpeciesAbundanceDietaryRangeRatio}
    \frac{A_i}{w^i}=\left(w_2-w_1\right)\chom.
\end{equation}
Without this, \((c*g)(x)\) would not be equal valued at all points along the resource axis (\(c\) would not be the preimage of the homogeneous steady state).  This condition means that \(A_i\propto w^i\).  Thus, larger \(w^i\) (the dietary range of species \(i\)) gives a larger species population size, making that species more resistant to stochastic extinction.  From this point of view, the most stable consumer population structure is one with a single group of ``generalists'', whose dietary range spans the entire resource space (\(w=1\)).  The effect of this has been noted in stochastic simulations.  Whilst some of the lower \(n\) patterns (Figure~\ref{fig:DominantEigenmodeFory2VSy1}) can be maintained for long periods of time in stochastic simulations, higher \(n\) patterns, or patterns with overlap of the dietary ranges of different species, are far less stable and prove short lived in stochastic simulations.  The larger the system however, the more likely it is that such patterns can be stable, with large enough numbers of each species to avoid stochastic extinction for longer periods of time.

\begin{figure}[htb!]
    \centering
    \includegraphics[width=\linewidth]{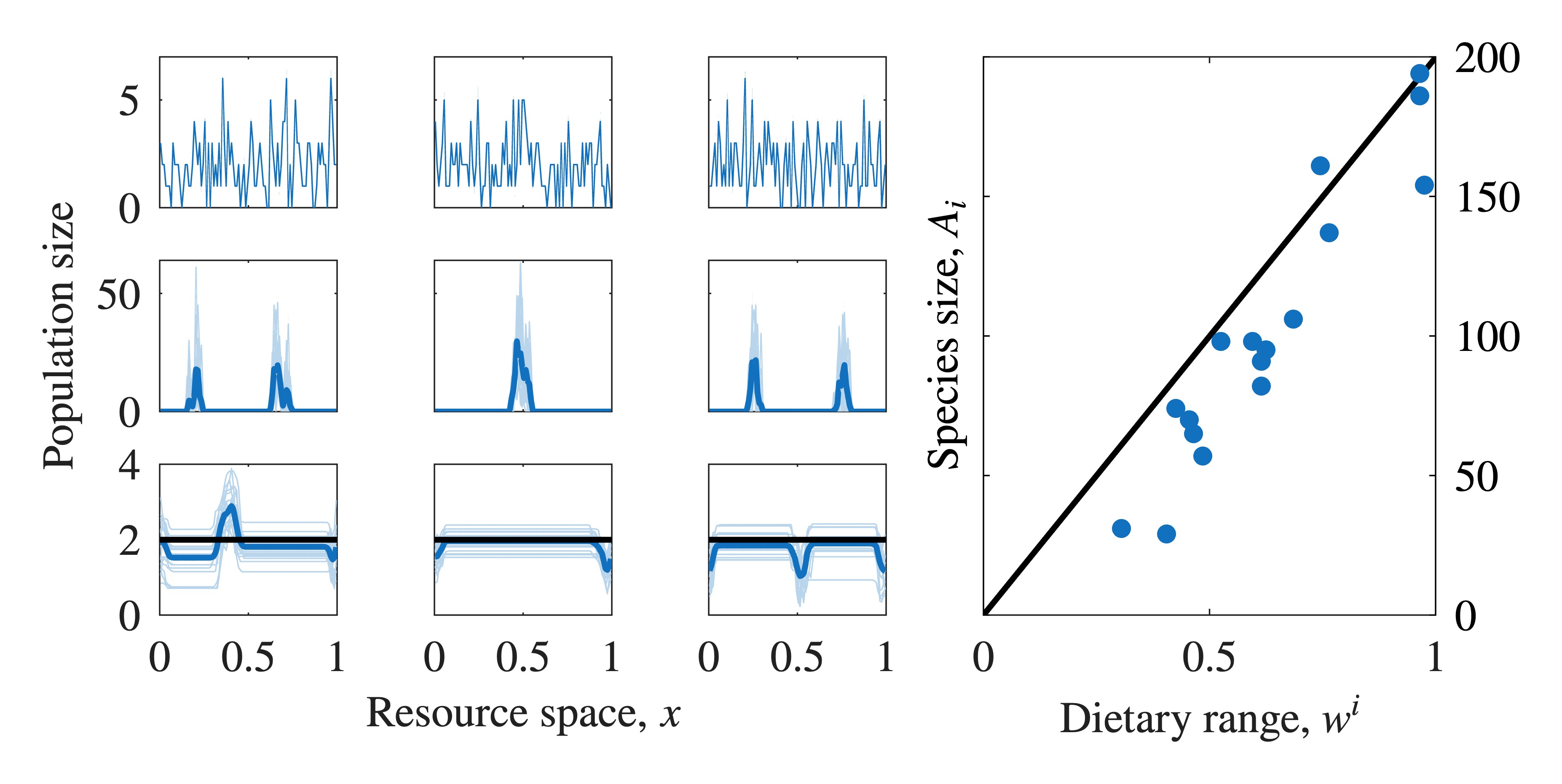}
    \caption{The three left hand columns display consumer populations from three distinct simulations up to \(t=200\).  Population numbers are summed over dietary range.  The top row shows the initial population and the middle row shows the final 20 snapshots of the simulated population in light blue, with the average in thick dark blue.  The bottom row shows the same data as the middle row, but each consumer distribution has been convolved with the resource preference distribution.  The theoretical prediction of the convolved distribution described by Equation~\eqref{eq:FlatConvolutionCondition} is shown in black.  The right hand plot shows the population size and dietary range of each species in 10 separate simulations at \(t=200\).  The black line shows the theoretical ratio of \(A_i\) to \(w^i\) given in Equation~\eqref{eq:SpeciesAbundanceDietaryRangeRatio}.  \(A_i\) and \(w^i\) were determined from simulation data using a clustering algorithm based on \(k\)-medoids.  The parameters of these systems were \(\alpha=2\times10^4\), \(\beta=1\), \(\delta=2\), \(\gamma=0.01\), \(\kappa=1\times10^5\), \(\mu_x=\mu_w=10^{-5}\), \(L=1\) and \(0\leq w\leq1\).  Data were recorded at 100 evenly spaced time points.  Resource-dietary range space was discretised into a \(100\times100\) grid.}
    \label{fig:Summary}
\end{figure}

Figure~\ref{fig:Summary} shows examples of consumer populations evolving towards distributions that satisfy the preimage condition.  The initial conditions in each simulation were random: consumers were placed uniformly at random within resource-dietary range space; the initial population size was equivalent to that of the homogeneous steady state consumer population.  The various population structures shown around the end of the simulations at \(t=200\) were contingent on the historic timing of stochastic events (path dependent consumer distributions).  Column one shows two species with dietary ranges \(w=0.455\) and \(w=0.625\); column two shows a single species with dietary range \(w=0.975\); column three shows two species with more similar dietary ranges, \(w=0.425\) and \(w=0.525\).  Despite this variety of evolved consumer populations, each approximately lies within the quasi-steady manifold, demonstrated by the convolution of each with the resource preference distribution fluctuating about the theoretical prediction given by the preimage condition, Equation~\eqref{eq:FlatConvolutionCondition}.  Thus, each population evolves towards the manifold, even if the details of that evolutionary trajectory can vary significantly.  In addition, the right hand plot in Figure~\ref{fig:Summary} shows that the individual species that form in each simulation obey Equation~\eqref{eq:SpeciesAbundanceDietaryRangeRatio} approximately.  Despite high levels of noise in the population sizes, the theoretical predictions match well with the simulated populations, and demonstrate that although the exact consumer population structure cannot be predicted, each evolved population is characterised by lying within the quasi-steady manifold.

\section{Discussion}
In this study, we analysed a continuous space model of proliferating resources and consumers which depended upon those resources to survive.  We showed how the deterministic equations have steady states which are homogeneous only, but that demographic noise can maintain regularly spaced species along the resource axis.  We then demonstrated how the number of such species could be predicted by analysing the equations linearised about the non-extinction homogeneous steady state.  This model was extended to allow to vary the dietary range of each consumer's resource preference distribution, as well as the resources contained within the dietary range.  This led to a far greater variety of possible states the system could exhibit, including a manifold onto which the system collapsed rapidly, but within which it evolved slowly.  The range of states maintained by demographic noise was shown to be far greater in this extended model, meaning the patterns of resource use seen were far more varied than for the fixed dietary range model, as well as being highly dependent on initial conditions and stochastic fluctuations.  Analysis of the system also showed that as the range of permitted dietary ranges was reduced, the predictability of the system's state increased until, in the limit, the original model was recovered.  Due to greater restrictions on the population size of species with narrower dietary ranges, such populations were more at risk of stochastic extinction, making generalists strategies the most long lived and stable.

The findings presented here are highly theoretical and our model describes a null case such that we can investigate the behaviour of a system of resources and consumers given no complicating features.  Any realistic behaviours not displayed by the model could suggest further mechanisms acting to produce those behaviours.  Despite its parsimonious nature, the model does display behaviours similar to that described empirically for real systems.  Viewing dietary range as a narrow definition of niche breadth \citep{Carscadden2020,Sexton2017} allows us to compare our results with other theoretical and empirical studies on this general theme.  Firstly, the idea that generalist consumers have larger population sizes has some empirical support \citep{Quintero2024}.  Similarly for the idea that generalist diets will evolve at the expense of greater numbers of species \citep{Granot2020}, as available niche space is ``taken up'' by species with greater dietary ranges.  However, there is inconsistent evidence on links between niche breadth and species coexistence \citep{Carscadden2020}.

The role of environmental heterogeneity is not addressed in this study.  It is interesting to note, however, that homogeneous environments, as studied here, are usually found to promote specialists \citep{Sexton2017,Kassen2002}.  Although possible that our homogeneous environment may slow the general evolutionary tendency towards generalism compared to a heterogeneous environment, it is apparent that some extension to the model may be required to prevent this macroevolutionary trend.  A trade-off making generalist diets less favourable is a clear route to achieving this.  There is, however, little support for the existence of such a trade-off \citep{Bebber2022,Sexton2017}.  As such, another consideration may be required to reproduce the observation of persistent specialist strategies.  Another possibility is a tuning of parameters concerning evolvability.  The comparisons in Figure~\ref{fig:EvolvePositionVSEvolveWidthFollowingExtinction} shows a possibility for maintaining specialist species.  Having position evolve more rapidly than dietary range induces speciation rather than dietary range expansion following an extinction event, preventing an increase in the average value of \(A_i\).  The is evidence of differential rates of evolution with niche breadth, with specialists having higher diversification rates \citep{Qiao2016}.  If speciation rates were higher for specialists then the effect just described may be enhanced.  We note however that support for correlations between niche breadth and diversification rates is mixed \citep{Sexton2017}.

It would be relatively simple to add temporal heterogeneity into our model or, indeed, heterogeneity in various parameters in resource space, \(x\) or dietary range, \(w\), at least in simulations.  However, adding in spatial structure would require more work.  This would be most easily implemented as extra dimensions in which the consumer population and resources live.  Adding additional spatial dimensions would allow consumers to feed within spatial ranges that could coevolve with position in resource space and dietary range.  Heterogeneous resource distributions would impact which spatial regions each species could survive within.  Spatial heterogeneity, although thought to be less important than temporal heterogeneity for the evolution of niche breadth \citep{Lin2017,Sexton2017}, has been shown to interact with niche breadth to affect the spatial structure of parameters such as abundances and the geographical range of species \citep{Carscadden2020}.  As studies generally find that greater movement favours a broader niche \citep{Sexton2017}, species with larger dietary ranges may be able to colonise greater geographical ranges, potentially amplifying the trend towards generalism.

Despite our interest in the development of a model that allows for the evolution of specialist and generalist species which occupy the same ecosystem, there are also empirical studies which point to a remarkable universality of dietary diversity \citep{Hutchinson2022,Rossberg2010}.  These findings indicate that there is some, as yet unknown, mechanism which can generate a sharp constraint on dietary diversity.  The shape of the dietary preference distribution observed in these studies can be implemented as a non-uniform \(g(x)\) in our model (stochastic simulations of this scenario are not dissimilar to results presented here for uniform \(g(x)\)), but in the extended model with dietary range evolution, the trend towards generalism will still be observed.  Thus, it is an interesting direction for future research to ask what additions to the model could reproduce this observation of stability at some intermediate value of dietary range.

\section*{Conflict of interest}
The authors have no competing interests to declare that are relevant to the content of this article.

\section*{Data availability}
There is no data associated with this manuscript.

\section*{Code availability}
Code for producing the figures in the manuscript, performing stochastic simulations of the model with variable dietary range (note that the model without variable dietary range can be simulated by adjusting code parameters) and performing matrix analysis can be found in a GitHub repository: \url{https://github.com/rEactorMB/SlowEvolutionTowardsGeneralismInAModelOfVariableDietaryRange.git}.  This repository also contains the data to produce the figures in the manuscript.

\section*{Acknowledgement}
We thank Matthew Wills for his extensive insights into the biological framing of the mathematical results presented here.

We would also like to thank Axel Rossberg for insightful discussions and for pointing out some relevant and intriguing empirical work.

EB would like to acknowledge EPSRC DTP funding [EP/W524712/1].

\appendix
\section{Appendix}
\subsection{Deriving the deterministic equations}\label{sec:DerivationOfPDEs}
To derive the partial differential equations, we begin with a description of the microscopic interactions within and between resources and consumers.  The state of the system at any time can be defined by the following two distributions:
\begin{align*}
    r(t,x)&=\frac{1}{R}\sum_{i=1}^{N(t)}\delta\left(x-x^r_i(t)\right),\\
    c(t,x)&=\frac{1}{C}\sum_{i=1}^{M(t)}\delta\left(x-x^c_i(t)\right).
\end{align*}
There are \(N(t)\) resources in the system and \(M(t)\) consumers.  \(R\) and \(C\) are large parameters related to the number of resources and consumers in the system respectively.  The \(i\)th resource has position \(x^r_i\) in resource space and the \(i\)th consumer has position \(x^c_i\).  We wish to describe the evolution of the system based on the rules described above.  To do so, we define \(P(t,r(t,x),c(t,x))\equiv P(r,c)\) as the probability for the distribution of resources and consumers to be \(r(t,x)\) and \(c(t,x)\) at time \(t\) respectively.  To change the state of the system, we can add and remove resources and consumers.  To this end, we define the following operators \citep{Rogers2012}:
\begin{align*}
    \Delta^\pm_{r,y}F[r,c]&=F\left[r\pm\frac{1}{R}\delta(x-y),c\right],\\
    \Delta^\pm_{c,y}F[r,c]&=F\left[r,c\pm\frac{1}{C}\delta(x-y)\right].
\end{align*}
Thus, we can write the master equation for this model as
\begin{multline*}
    \pdiff{}{t}P(r,c)=R\alpha\int\left(\Delta^-_{r,y}-1\right)\left[P(r,c)\right]\dd{y}+\\
    R\frac{\alpha}{\kappa}\int\left(\Delta^+_{r,y}-1\right)\left[rP(r,c)\right]\dd{y}+R\int\left(\Delta^+_{r,y}-1\right)\left[(c*g)rP(r,c)\right]\dd{y}+\\
    C\beta\int\left(\Delta^-_{c,y}-1\right)\left[(c*\Phi)P(r,c)\right]\dd{y}+C\delta\int\left(\Delta^+_{c,y}-1\right)\left[cP(r,c)\right]\dd{y}+\\
    C\gamma\int\left(\Delta^-_{c,y}-1\right)\left[(r*g)cP(r,c)\right]\dd{y}.
\end{multline*}
The first integral accounts for the production of new resources, the second for the removal of resources due to intraspecific competition (preventing proliferation past carrying capacity \(\kappa\)), the third for the consumption of resources by the consumers (\(g(x)\) is a resource preference distribution representing the resource preference of each consumer which we normalise as a probability distribution).  The fourth integral accounts for birth of consumers with mutation (\(\Phi(x)\) is the Gaussian distribution with variance \(2\mu\)), the fifth accounts for the death of consumers and the final integral accounts for the increase in birth rate of consumers as they use available resources.  We can make progress by considering the following expansion,
\begin{align*}
    \Delta^\pm_{r,y}F[r,c]&=F\left[r\pm\frac{1}{R}\delta(x-y),c\right],\\
    &=F\left[r,c\right]\pm\frac{1}{R}\fdiff{}{r}F[r,c]+\mathcal{O}\left((R)^{-2}\right).
\end{align*}
Thus,
\begin{equation*}
    \Delta^\pm_{r,y}=1\pm\frac{1}{R}\fdiff{}{r}+\mathcal{O}\left((R)^{-2}\right),
\end{equation*}
and similarly
\begin{equation*}
    \Delta^\pm_{c,y}=1\pm\frac{1}{C}\fdiff{}{c}+\mathcal{O}\left((C)^{-2}\right).
\end{equation*}
Putting this into our master equation and keeping only the terms \(\mathcal{O}\left((R)^0\right)\) and \(\mathcal{O}\left((C)^0\right)\) (i.e., assuming a large system such that demographic effects are lost) gives
\begin{multline*}
    \pdiff{}{t}P(r,c)=-\alpha\int\fdiff{}{r}\left[P(r,c)\right]\dd{y}+\\
    \frac{\alpha}{\kappa}\int\fdiff{}{r}\left[rP(r,c)\right]\dd{y}+\int\fdiff{}{r}\left[(c*g)rP(r,c)\right]\dd{y}-\\
    \beta\int\fdiff{}{c}\left[(c*\Phi)P(r,c)\right]\dd{y}+\delta\int\fdiff{}{c}\left[cP(r,c)\right]\dd{y}-\\
    \gamma\int\fdiff{}{c}\left[(r*g)P(r,c)\right]\dd{y}.
\end{multline*}
Combining the integrals gives
\begin{multline*}
    \pdiff{}{t}P(r,c)=-\int\fdiff{}{r}\left[\left(\alpha-\frac{\alpha}{\kappa}r-(c*g)r\right)P(r,c)\right]\dd{y}-\\
    \int\fdiff{}{c}\left[\left(\beta(c*\Phi)-\delta c+\gamma(r*g)\right)P(r,c)\right]\dd{y}.
\end{multline*}
We now define
\begin{align*}
    \mathcal{R}&=\alpha-\frac{\alpha}{\kappa}r-(c*g)r,\\
    \mathcal{C}&=\beta(c*\Phi)-\delta c+\gamma(r*g)c,
\end{align*}
such that \(r\) and \(c\) are the solutions to
\begin{align}
    \pdiff{r}{t}&=\mathcal{R},\label{eq:partial_r_partial_t}\\
    \pdiff{c}{t}&=\mathcal{C}.\label{eq:partial_c_partial_t}
\end{align}
Given some initial condition, \(r\) and \(c\) follow the trajectories \(r^*\) and \(c^*\) which are the solutions to \(\pdiff{r^*}{t}=\mathcal{R}^*\) and \(\pdiff{c^*}{t}=\mathcal{C}^*\) respectively and \(\mathcal{R}^*\equiv\mathcal{R}(r^*,c^*)\), \(\mathcal{C}^*\equiv\mathcal{C}(r^*,c^*)\).  We now state that \(P(r,c)=\delta(r-r^*)\delta(c-c^*)\), which means that
\begin{multline*}
    \pdiff{}{t}\left[\delta(r-r^*)\delta(c-c^*)\right]=\\
    -\int\fdiff{}{r}\left[\mathcal{R}\delta(r-r^*)\delta(c-c^*)\right]\dd{y}-\int\fdiff{}{c}\left[\mathcal{C}\delta(r-r^*)\delta(c-c^*)\right]\dd{y},
\end{multline*}
which becomes
\begin{multline}\label{eq:FinalLiouvilleEquation}
    \pdiff{}{t}\left[\delta(r-r^*)\delta(c-c^*)\right]=\\
    -\int\mathcal{R}^*\fdiff{}{r}\left[\delta(r-r^*)\right]\delta(c-c^*)\dd{y}-\int\mathcal{C}^*\delta(r-r^*)\fdiff{}{c}\left[\delta(c-c^*)\right]\dd{y}.
\end{multline}
We also see that
\begin{multline*}
    \pdiff{}{t}\left[\delta(r-r^*)\delta(c-c^*)\right]=\\
    \int\left\{-\fdiff{}{r^*}\left[\delta(r-r^*)\delta(c-c^*)\right]\pdiff{r^*}{t}-\fdiff{}{c^*}\left[\delta(r-r^*)\delta(c-c^*)\right]\pdiff{c^*}{t}\right\}\dd{y},
\end{multline*}
which becomes, using Equations \eqref{eq:partial_r_partial_t} and \eqref{eq:partial_c_partial_t},
\begin{multline}\label{eq:DerivativeOfProbabilityDeltaFunction}
    \pdiff{}{t}\left[\delta(r-r^*)\delta(c-c^*)\right]=\\
    -\int\fdiff{}{r}\left[\delta(r-r^*)\right]\delta(c-c^*)\mathcal{R}^*-\int\delta(r-r^*)\fdiff{}{c^*}\left[\delta(c-c^*)\right]\mathcal{C}^*\dd{y}.
\end{multline}
Thus, since the right hand side of Equation~\eqref{eq:FinalLiouvilleEquation} and Equation~\eqref{eq:DerivativeOfProbabilityDeltaFunction} are equal, we see that Equations \eqref{eq:partial_r_partial_t} and \eqref{eq:partial_c_partial_t} hold.

We can rewrite the term including the Gaussian distribution as follows.  Since
\begin{equation*}
    \pdiff{\Phi}{\mu}=\pdiffn{\Phi}{x}{2},
\end{equation*}
as \(\mu\) becomes very small, we can make the approximation
\begin{equation*}
    \int c(y)\Phi(x-y)\dd{y}\approx\int c(y)\delta(x-y)\dd{y}+\mu\int\pdiffn{c(y)}{y}{2}\delta(x-y)\dd{y}.
\end{equation*}
Thus, we can model the dynamics described above with the following partial differential equations:
\begin{align*}
    \frac{\partial r}{\partial t}&=\alpha\left(1-\frac{r}{\kappa}\right)-r(c*g),\\
    \frac{\partial c}{\partial t}&=(\beta-\delta)c+\gamma c(r*g)+\mu\nabla^2c.
\end{align*}

\subsection{Coarse-graining the modified equations}\label{sec:CoarseGrainingDietaryRange}
In order to make progress with analysis of the modified partial differential equations which allow for dietary range to evolve in addition to the resources included in a species' dietary range, we can coarse-grain the range of dietary range values.  This allows us to work with the vector quantity
\begin{equation*}
    \boldsymbol{\psi}=\begin{bmatrix}
        \tilde{R}_k\left(t\right)\\
        \tilde{C}_k\left(t,\vw\right)
    \end{bmatrix}=\begin{bmatrix}
        \tilde{R}_k\left(t\right)&
        \tilde{C}_k\left(t,w_{(1)}\right)&
        \tilde{C}_k\left(t,w_{(2)}\right)&
        \cdots&
        \tilde{C}_k\left(t,w_{(N)}\right)
    \end{bmatrix}^T
\end{equation*}
where \(w\) has been coarse-grained into \(N\) discrete values such that
\begin{equation*}
    \vw=\begin{bmatrix}
        w_{(1)}&w_{(2)}&\cdots&w_{(N)}
        \end{bmatrix}^T.
\end{equation*}
Equations \eqref{eq:tEvolutionRTilde_k} and \eqref{eq:tEvolutionCTilde_k} become, following the coarse-graining of \(w\) (and dropping \(t\) for clarity),
\begin{align*}
    \pdiff{\tilde{R}_k}{t}&=-\frac{\alpha}{\rhom}\tilde{R}_k-\rhom\sum_{n=1}^{N}\tilde{C}_{k,w_{(n)}}G_{k,w_{(n)}}W_w,\\
    \pdiff{\tilde{C}_{k,w_{(n)}}}{t}&=\left(\beta-\delta+\gamma\rhom\right)\tilde{C}_{k,w_{(n)}}+\gamma\chom\tilde{R}_kG_{k,w_{(n)}}-\\
    &\qquad\mu_xk^2\tilde{C}_{k,w_{(n)}}+\frac{\mu_w}{W_w^2}\boldsymbol{\Delta}_w\tilde{C}_{k,\vw}.
\end{align*}
By the definition of \(\rhom\) we see that the first term on the right hand side of the second equation is zero, meaning
\begin{equation*}
    \pdiff{\tilde{C}_{k,w_{(n)}}}{t}=\gamma\chom\tilde{R}_kG_{k,w_{(n)}}-\mu_xk^2\tilde{C}_{k,w_{(n)}}+\frac{\mu_w}{W_w^2}\boldsymbol{\Delta}_w\tilde{C}_{k,\vw}.
\end{equation*}
We have defined \(\tilde{C}_{k,w_{(n)}}=\tilde{C}_k(t,w_{(n)})\) as the \(n\)th term of
\begin{equation*}
    \tilde{C}_{k,\vw}=\begin{bmatrix}
        \tilde{C}_k\left(t,w_{(1)}\right)&\tilde{C}_k\left(t,w_{(2)}\right)&\cdots&\tilde{C}_k\left(t,w_{(N)}\right)
    \end{bmatrix}^T
\end{equation*}
and \(G_{k,w_{(n)}}=G_k(w_{(n)})\) is the \(n\)th term of \(G_{k,\vw}\) analogously.  \(\boldsymbol{\Delta}_w\) is the matrix representation of the Laplacian in \(w\) and \(W_w\) is the spacing between values in the vector \(\vw\) (\(W_w=w_{(2)}-w_{(1)}=\frac{w_2-w_1}{N}\)).  We can use these two equations to write the partial derivative of \(\boldsymbol{\psi}\) with respect to time as a matrix equation:
\begin{equation}\label{eq:MatrixADifferentialEquation}
    \pdiff{\boldsymbol{\psi}}{t}=\vA\boldsymbol{\psi},
\end{equation}
where \(\vA=\vA_R+\vA_C\) is defined as follows:
\begin{equation*}
    \vA_R=\begin{bmatrix}
        -\frac{\alpha}{\rhom}&-\rhom G_{k,\vw^T}W_w\\
        \mathbf{0}&\mathbf{0}
    \end{bmatrix},
\end{equation*}
\begin{equation*}
    \vA_C=\begin{bmatrix}
        0&\mathbf{0}\\
        \gamma\chom G_{k,\vw}&\mathbf{0}\\
    \end{bmatrix}-\mu_xk^2\begin{bmatrix}
        0&\mathbf{0}\\
        \mathbf{0}&\mathbb{1}_N
    \end{bmatrix}+\frac{\mu_w}{W_w^2}\begin{bmatrix}
        0&\mathbf{0}\\
        \mathbf{0}&\boldsymbol{\Delta}_w
    \end{bmatrix}
\end{equation*}
where \(\mathbb{1}_N\) is the \(N\times N\) identity matrix and
\begin{equation*}
    \boldsymbol{\Delta}_w=\begin{bmatrix}
        -1&1&0&\cdots&0&0\\
        1&-2&1&\cdots&0&0\\
        0&1&-2&\cdots&0&0\\
        \vdots&\vdots&\vdots&\ddots&\vdots&\vdots\\
        0&0&0&\cdots&-2&1\\
        0&0&0&\cdots&1&-1
    \end{bmatrix}
\end{equation*}
for reflecting boundaries at \(w_1\) and \(w_2\).  We can now find the eigenvalues and eigenvectors of the matrix \(\vA\).  These detail, for a given value of \(k\), which perturbed modes die away most slowly.

\subsection{The effect of non-periodic boundaries}\label{sec:NeumannBoundaries}
Throughout the paper, periodic boundary conditions are used to simplify analytical exploration of the solutions to the resource-consumer PDE system.  This is not a realistic modelling choice.   In reality, there would be an ``edge'' to the resources, limiting species whose lineages evolve preferences for those resources at the limits of the resource distribution.  Continued evolution towards this edge will result in a species having fewer and fewer resources it can access, rather than being able to access the resources at the opposite end of the distribution.  

\begin{figure}[htb!]
    \centering
    \includegraphics[width=\textwidth]{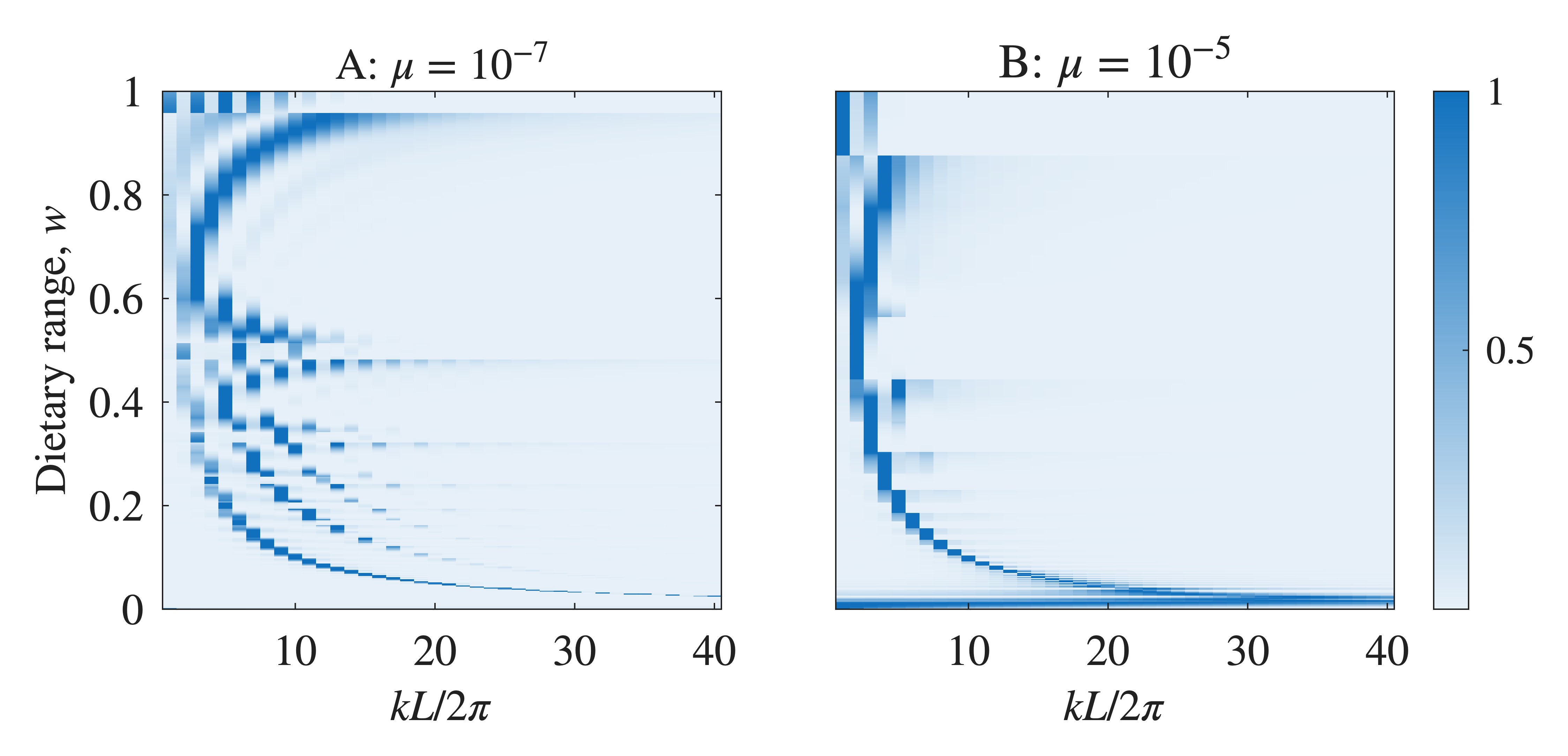}
    \caption{The Fourier spectrum of the leading eigenvector of the matrix describing the evolution of linear perturbations about the homogeneous steady state of the model without dietary range evolution for various values of \(\frac{kL}{2\pi}\) between zero and 40 and values for the dietary range, \(w\), between zero and one.  For each value of \(w\), the spectrum was normalised by dividing all values of the spectrum by the power of the highest peak.  The parameter values for the system were \(\alpha=2\times10^4\), \(\beta=1\), \(\delta=2\), \(\gamma=0.01\), \(\kappa=1\times10^5\) and \(L=1\).  In A, \(\mu=10^{-7}\) and in B, \(\mu=10^{-5}\).  A Fourier spectrum was generated for 500 values of \(w\), chosen uniformly between 0 and 1.  The matrix \(\mM\) was \(2000\times2000\) in each case (\(N=1000\)).}
    \label{fig:PowerSpectrum_Neumann}
\end{figure}

The similarity in the patterns formed when periodic boundaries are replaced with Neumann boundaries can be demonstrated by comparing Figure~\ref{fig:PowerSpectrum_0<w<1} to Figure~\ref{fig:PowerSpectrum_Neumann}.  Figure~\ref{fig:PowerSpectrum_Neumann} was generated by coarse-graining the model without dietary range evolution once it has been linearised about its homogeneous steady state.  Since we are working in a bounded domain with non-periodic boundary conditions, Fourier transforms are not useful for removing the convolution terms as they were in the periodic domain.  In order to remove these integral terms to analyse the dominant linear perturbations about the homogeneous steady state, we instead coarse-grain the resource space domain, converting integrals into matrix products.  Thus, the perturbations \(\tilde{r}(t,x)\) and \(\tilde{c}(t,x)\), whose time evolution is described by the equations
\begin{align*}
    \pdiff{\tilde{r}}{t}&=-\frac{\alpha\tilde{r}}{\kappa}-\tilde{r}\chom-\rhom(\tilde{c}*g),\\
    \pdiff{\tilde{c}}{t}&=\gamma\chom\tilde{r}*g+\mu\nabla^2\tilde{c},
\end{align*}
become length \(N\) vectors \(\tilde{\vr}\) and \(\tilde{\vc}\).  We now can write the two evolution equations as a single matrix equation of the form
\begin{equation*}
    \pdiff{}{t}\begin{bmatrix}
        \tilde{\vr}\\
        \tilde{\vc}
    \end{bmatrix}=\mM\begin{bmatrix}
        \tilde{\vr}\\
        \tilde{\vc}
    \end{bmatrix}.
\end{equation*}
The matrix \(\mM\) has block structure
\begin{equation*}
    \mM=\begin{bmatrix}
        \mA&\mB\\
        \mC&\mD
    \end{bmatrix}.
\end{equation*}
We can perform the matrix product of \(\mM\) and \(\begin{bmatrix}\tilde{\vr}&\tilde{\vc}\end{bmatrix}^T\) and compare the terms to those in the evolution equations for \(\tilde{r}(t,x)\) and \(\tilde{c}(t,x)\).  Thus, we see that the four blocks of \(\mM\) are defined as follows:
\begin{align*}
    \mA&=\left(-\frac{\alpha}{\kappa}-\chom\right)\mathbb{1}_N,\\
    \mB&=-\rhom\mG,\\
    \mC&=\gamma\chom\mG,\\
    \mD&=\frac{\mu}{N^2}\boldsymbol{\Delta}.
\end{align*}
\(\mathbb{1}_N\) is the \(N\times N\) identity matrix.  \(\boldsymbol{\Delta}\) is defined identically to \(\boldsymbol{\Delta_w}\) (see Section~\ref{sec:CoarseGrainingDietaryRange}).  The \(N\times N\) matrix \(\mG\) is the matrix representation of the (bounded uniform) resource preference distributions for non-periodic boundaries on resource space.  For dietary range \(w\) it has values \(\frac{1}{w}\) in a diagonal strip and zeros everywhere else.  The maximum number of non-zero values in a row is \(wN\) (rounded), but this is reduced for rows representing individuals that are close enough to the boundaries on resource space for their preference distribution to be truncated.

To find the modes that die away most slowly within this system, indicative of the number of coexisting species for a given value of dietary range predicted by this model, we analyse the eigenvalues of the matrix \(\mM\).  The eigenvalue with maximum real part was found and its corresponding eigenvector extracted.  A fast Fourier transform was then taken of the second half of this eigenvector (the part corresponding to the consumer distribution perturbation).  This process is repeated for each value of \(w\).  Combining these spectra and normalising each one such that the maximum power was 1 in each results in the plot shown in Figure~\ref{fig:PowerSpectrum_Neumann}.

The most significant difference seen between the two versions of the spectrum, with and without periodic boundaries, is the loss of structure for \(w>0.5\) in the non-periodic case.  This means that in the non-periodic case, patterns where many species compete for each resource are more strongly damped than in the periodic case.  This behaviour is more similar to what is observed in stochastic simulations, even with periodic boundary conditions.  The similarity in appearance of Figure~\ref{fig:PowerSpectrum_0<w<1} and Figure~\ref{fig:PowerSpectrum_Neumann} demonstrate that the results found for periodic boundaries should hold in general for a more realistic system which does not have a periodic resource domain.

\begin{figure}[htb!]
    \centering
    \includegraphics[width=\linewidth]{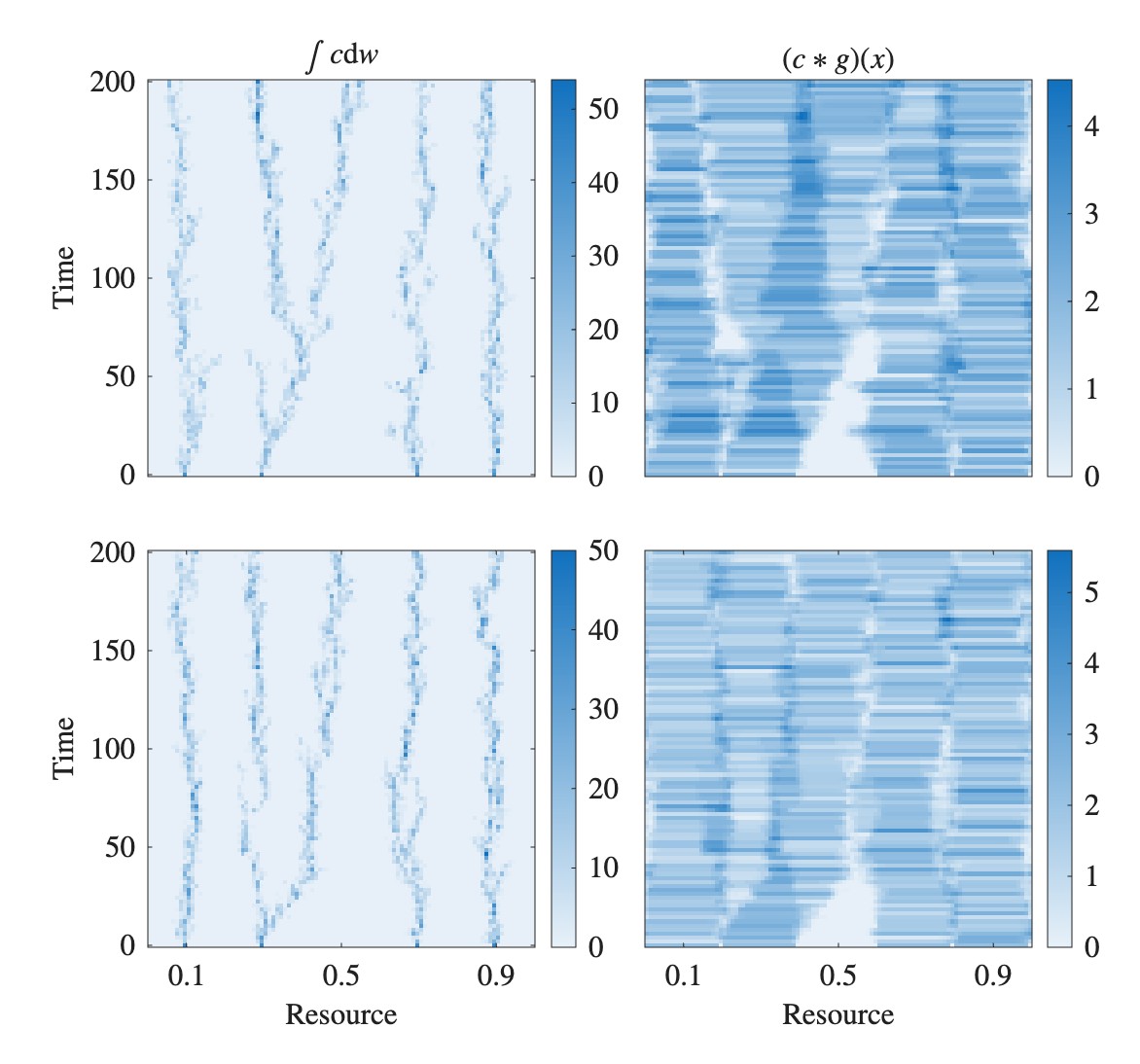}
    \caption{The system in both rows is in the same state at \(t=0\): five species with dietary range \(w=0.2\), equally spaced along the resource axis, \(x\) (100 bins), at the moment the centrally positioned species goes extinct.  The upper panels show the evolution of this system when \(0<w<1\) (50 bins), and the lower panels show the evolution when \(0.19<w<0.21\) (5 bins): an unrestricted and a restricted evolution of dietary range scenario.  At the end of each simulation,  Equation~\eqref{eq:FlatConvolutionCondition} is satisfied with five species.  In the unrestricted case, these species are highly heterogeneous in their dietary range (note the different widths of the ``columns'' in the right hand plot), whereas the restricted case results in the evolution of five approximately equal species.  The boundary conditions on resource space, \(x\), were Neumann in this simulation.  The parameters of these systems were \(\alpha=2\times10^4\), \(\beta=1\), \(\delta=2\), \(\gamma=0.01\), \(\kappa=1\times10^5\), \(\mu_x=\mu_w=10^{-5}\) and \(L=1\).}
    \label{fig:SimulationResultsNeumann}
\end{figure}

To investigate the effect this boundary alteration has on the behaviour of the model with dietary range evolution, we can run simulations of the model and compare the results to those for a system with with periodic boundaries.  Figure~\ref{fig:SimulationResultsNeumann} shows the results which are directly comparable to those in Figure~\ref{fig:FullVsRestricted_y_FollowingExtinction}.  The only difference in the system set up is the boundary conditions imposed on resource space.  We see that the difference is minimal, with the main effect being to ``pin'' the pattern to a central location between the two boundaries.  This is a result of the unfavourable evolution towards the limits of resources: close to the boundary (with close defined relative to the dietary range of a species), resource preference distributions are truncated, reducing the potential for a species to access food and so reducing the stable population size of that species.  This position-dependent effect whereby evolution towards the boundary is detrimental, but evolution away from it is beneficial, means that the group of species represented by a single pattern is most stable when no species sits too close to the boundary.  This is not the case with periodic boundaries where the stability of the pattern is not affected by rotating each species around the cylindrical resource space.

\printbibliography

@article{ButlerGoldenfeld2009,
  title = {Robust ecological pattern formation induced by demographic noise},
  author = {Butler, Thomas and Goldenfeld, Nigel},
  journal = {Phys. Rev. E},
  volume = {80},
  issue = {3},
  pages = {030902},
  numpages = {4},
  year = {2009},
  month = {09},
  publisher = {American Physical Society},
  doi = {10.1103/PhysRevE.80.030902},
  url = {https://link.aps.org/doi/10.1103/PhysRevE.80.030902}
}

@article{Butler2011,
  title = {Fluctuation-driven Turing patterns},
  author = {Butler, Thomas and Goldenfeld, Nigel},
  journal = {Phys. Rev. E},
  volume = {84},
  issue = {1},
  pages = {011112},
  numpages = {12},
  year = {2011},
  month = {07},
  publisher = {American Physical Society},
  doi = {10.1103/PhysRevE.84.011112},
  url = {https://link.aps.org/doi/10.1103/PhysRevE.84.011112}
}

@article{Leimar2013,
title = {Limiting similarity, species packing, and the shape of competition kernels},
journal = {Journal of Theoretical Biology},
volume = {339},
pages = {3-13},
year = {2013},
note = {Peter Abrams Honor},
issn = {0022-5193},
doi = {10.1016/j.jtbi.2013.08.005},
url = {https://www.sciencedirect.com/science/article/pii/S0022519313003779},
author = {Olof Leimar and Akira Sasaki and Michael Doebeli and Ulf Dieckmann}
}

@article{Pigolotti2007,
  title = {Species Clustering in Competitive Lotka-Volterra Models},
  author = {Pigolotti, Simone and López, Cristóbal and Hernández-García, Emilio},
  journal = {Phys. Rev. Lett.},
  volume = {98},
  issue = {25},
  pages = {258101},
  numpages = {4},
  year = {2007},
  month = {06},
  publisher = {American Physical Society},
  doi = {10.1103/PhysRevLett.98.258101},
  url = {https://link.aps.org/doi/10.1103/PhysRevLett.98.258101}
}

@article{Pigolotti2010,
    author = {Pigolotti, S. and López, C. and Hernández-García and Andersen, K. H.},
    title = {How Gaussian competition leads to lumpy or uniform species distributions},
    journal = {Theoretical Ecology},
    year = {2010},
    volume = {3},
    pages = {89--96},
    doi = {10.1007/s12080-009-0056-2}
}

@article{HernandezGarcia2009,
author = {Hernández-García, Emilio  and López, Cristóbal  and Pigolotti, Simone  and Andersen, Ken H.},
title = {Species competition: coexistence, exclusion and clustering},
journal = {Philosophical Transactions of the Royal Society A: Mathematical, Physical and Engineering Sciences},
volume = {367},
number = {1901},
pages = {3183-3195},
year = {2009},
doi = {10.1098/rsta.2009.0086},
URL = {https://royalsocietypublishing.org/doi/abs/10.1098/rsta.2009.0086},
eprint = {https://royalsocietypublishing.org/doi/pdf/10.1098/rsta.2009.0086}
}

@article{Rogers2012,
doi = {10.1209/0295-5075/97/40008},
url = {https://dx.doi.org/10.1209/0295-5075/97/40008},
year = {2012},
month = {02},
publisher = {},
volume = {97},
number = {4},
pages = {40008},
author = {T. Rogers and A. J. McKane and A. G. Rossberg},
title = {Demographic noise can lead to the spontaneous formation of species},
journal = {Europhysics Letters}
}

@article{Rogers2015,
  title = {Modes of competition and the fitness of evolved populations},
  author = {Rogers, Tim and McKane, Alan J.},
  journal = {Phys. Rev. E},
  volume = {92},
  issue = {3},
  pages = {032708},
  numpages = {10},
  year = {2015},
  month = {09},
  publisher = {American Physical Society},
  doi = {10.1103/PhysRevE.92.032708},
  url = {https://link.aps.org/doi/10.1103/PhysRevE.92.032708}
}

@book{Levins1968,
 ISBN = {9780691079592},
 URL = {http://www.jstor.org/stable/j.ctvx5wbbh},
 author = {Richard Levins},
 publisher = {Princeton University Press},
 title = {Evolution in Changing Environments: Some Theoretical Explorations. (MPB-2)},
 urldate = {2025-12-03},
 year = {1968}
}

@book{Roughgarden1979,
  title={Theory of Population Genetics and Evolutionary Ecology: An Introduction},
  author={Roughgarden, J.},
  year={1979},
  publisher={Macmillan, London}
}

@article{MacArthur1967,
 ISSN = {00030147, 15375323},
 URL = {http://www.jstor.org/stable/2459090},
 author = {Robert MacArthur and Richard Levins},
 journal = {The American Naturalist},
 number = {921},
 pages = {377--385},
 publisher = {[The University of Chicago Press, The American Society of Naturalists]},
 title = {The Limiting Similarity, Convergence, and Divergence of Coexisting Species},
 urldate = {2025-12-03},
 volume = {101},
 year = {1967}
}

@article{Dehling2021,
author = {Dehling, D. Matthias and Bender, Irene M. A. and Blendinger, Pedro G. and Böhning-Gaese, Katrin and Muñoz, Marcia C. and Neuschulz, Eike L. and Quiti\'{a}n, Marta and Saavedra, Francisco and Santill\'{a}n, Vinicio and Schleuning, Matthias and Stouffer, Daniel B.},
title = {Specialists and generalists fulfil important and complementary functional roles in ecological processes},
journal = {Functional Ecology},
volume = {35},
number = {8},
pages = {1810-1821},
doi = {10.1111/1365-2435.13815},
url = {https://besjournals.onlinelibrary.wiley.com/doi/abs/10.1111/1365-2435.13815},
eprint = {https://besjournals.onlinelibrary.wiley.com/doi/pdf/10.1111/1365-2435.13815},
year = {2021}
}

@article{vonMeijenfeldt2023,
    author = {von Meijenfeldt, F.A.B. and Hogeweg, P. and Dutilh, B.E.},
    title = {A social niche breadth score reveals niche range strategies of generalists and specialists},
    journal = {Nat Ecol Evol},
    volume = {7},
    year = {2023},
    pages = {768--781},
    doi = {10.1038/s41559-023-02027-7}
}

@article{Grant2006,
author = {Peter R. Grant  and B. Rosemary Grant},
title = {Evolution of Character Displacement in Darwin's Finches},
journal = {Science},
volume = {313},
number = {5784},
pages = {224-226},
year = {2006},
doi = {10.1126/science.1128374},
URL = {https://www.science.org/doi/abs/10.1126/science.1128374},
eprint = {https://www.science.org/doi/pdf/10.1126/science.1128374}
}

@book{Darwin1859,
    author = {Darwin, C.},
    title = {On the origin of species},
    publisher = {J. Murrary},
    year = {1859}
}

@article{Dennis2011,
    author = {Dennis, Roger L. H. and Dapporto, Leonardo and Fattorini, Simone and Cook, Laurence M.},
    title = {The generalism–specialism debate: the role of generalists in the life and death of species},
    journal = {Biological Journal of the Linnean Society},
    volume = {104},
    number = {4},
    pages = {725-737},
    year = {2011},
    month = {10},
    issn = {0024-4066},
    doi = {10.1111/j.1095-8312.2011.01789.x},
    url = {https://doi.org/10.1111/j.1095-8312.2011.01789.x},
    eprint = {https://academic.oup.com/biolinnean/article-pdf/104/4/725/16707219/j.1095-8312.2011.01789.x.pdf},
}

@article{Kassen2002,
author = {Kassen, R.},
title = {The experimental evolution of specialists, generalists, and the maintenance of diversity},
journal = {Journal of Evolutionary Biology},
volume = {15},
number = {2},
pages = {173-190},
doi = {10.1046/j.1420-9101.2002.00377.x},
url = {https://onlinelibrary.wiley.com/doi/abs/10.1046/j.1420-9101.2002.00377.x},
eprint = {https://onlinelibrary.wiley.com/doi/pdf/10.1046/j.1420-9101.2002.00377.x},
year = {2002}
}

@article{Szabo2006,
author = {Szabó, Péter and Meszéna, Géza},
title = {Limiting similarity revisited},
journal = {Oikos},
volume = {112},
number = {3},
pages = {612-619},
doi = {10.1111/j.0030-1299.2006.14128.x},
url = {https://nsojournals.onlinelibrary.wiley.com/doi/abs/10.1111/j.0030-1299.2006.14128.x},
eprint = {https://nsojournals.onlinelibrary.wiley.com/doi/pdf/10.1111/j.0030-1299.2006.14128.x},
year = {2006}
}

@article{Barabas2009,
title = {When the exception becomes the rule: The disappearance of limiting similarity in the Lotka-Volterra model},
journal = {Journal of Theoretical Biology},
volume = {258},
number = {1},
pages = {89-94},
year = {2009},
issn = {0022-5193},
doi = {10.1016/j.jtbi.2008.12.033},
url = {https://www.sciencedirect.com/science/article/pii/S0022519309000034},
author = {György Barabás and Géza Meszéna}
}

@article{Letten2017,
author = {Letten, Andrew D. and Ke, Po-Ju and Fukami, Tadashi},
title = {Linking modern coexistence theory and contemporary niche theory},
journal = {Ecological Monographs},
volume = {87},
number = {2},
pages = {161-177},
doi = {10.1002/ecm.1242},
url = {https://esajournals.onlinelibrary.wiley.com/doi/abs/10.1002/ecm.1242},
eprint = {https://esajournals.onlinelibrary.wiley.com/doi/pdf/10.1002/ecm.1242},
year = {2017}
}

@article{Vamosi2014,
author = {Vamosi, Jana C.  and Armbruster, W. Scott  and Renner, Susanne S.},
title = {Evolutionary ecology of specialization: insights from phylogenetic analysis},
journal = {Proceedings of the Royal Society B: Biological Sciences},
volume = {281},
number = {1795},
pages = {20142004},
year = {2014},
doi = {10.1098/rspb.2014.2004},
URL = {https://royalsocietypublishing.org/doi/abs/10.1098/rspb.2014.2004},
eprint = {https://royalsocietypublishing.org/doi/pdf/10.1098/rspb.2014.2004}
}

@book{Lack1947,
  title={Darwin's finches},
  author={Lack, David},
  year={1947},
  publisher={Cambridge University Press}
}

@article{Bowman1961,
author="Bowman, R.I.",
title="Morphological differentiation and adaptation in the Galapagos finches",
journal="University of California publications in zoology",
year="1961",
volume="58",
pages="1-302",
URL="https://cir.nii.ac.jp/crid/1572261551130238976"
}

@book{Grant1999,
  title={Ecology and Evolution of Darwin's Finches},
  author={Grant, P. R.},
  year={1999},
  publisher={Princeton University Press}
}

@article{Foster2008,
  title={A geometric morphometric appraisal of beak shape in Darwin’s finches},
  author={Foster, D.J. and Podos, J. and Hendry, A.P.},
  journal={Journal of Evolutionary Biology},
  volume={21},
  number={1},
  pages={263--275},
  year={2008},
  publisher={Blackwell Publishing Ltd Oxford, UK},
  doi = {10.1111/j.1420-9101.2007.01449.x}
}

@article{Quintero2024,
author = {Castaño-Quintero, Sandra and Velasco, Julián and González-Voyer, Alejandro and Martínez-Meyer, Enrique and Yáñez-Arenas, Carlos},
title = {Niche position and niche breadth effects on population abundances: A case study of New World Warblers (Parulidae)},
journal = {Ecology and Evolution},
volume = {14},
number = {3},
pages = {e11108},
doi = {10.1002/ece3.11108},
url = {https://onlinelibrary.wiley.com/doi/abs/10.1002/ece3.11108},
eprint = {https://onlinelibrary.wiley.com/doi/pdf/10.1002/ece3.11108},
note = {e11108 ECE-2023-09-01713.R1},
year = {2024}
}

@article{Granot2020,
author = {Granot, Itai and Belmaker, Jonathan},
title = {Niche breadth and species richness: Correlation strength, scale and mechanisms},
journal = {Global Ecology and Biogeography},
volume = {29},
number = {1},
pages = {159-170},
doi = {10.1111/geb.13011},
url = {https://onlinelibrary.wiley.com/doi/abs/10.1111/geb.13011},
eprint = {https://onlinelibrary.wiley.com/doi/pdf/10.1111/geb.13011},
year = {2020}
}

@article{Carscadden2020,
author = {Carscadden, Kelly A. and Emery, Nancy C. and Arnillas, Carlos A. and Cadotte, Marc W. and Afkhami, Michelle E. and Gravel, Dominique and Livingstone, Stuart W. and Wiens, John J.},
title = {Niche Breadth: Causes and Consequences for Ecology, Evolution, and Conservation},
journal = {The Quarterly Review of Biology},
volume = {95},
number = {3},
pages = {179-214},
year = {2020},
doi = {10.1086/710388},
URL = {https://doi.org/10.1086/710388},
eprint = {https://doi.org/10.1086/710388}
}

@article{Sexton2017,
   author = "Sexton, Jason P. and Montiel, Jorge and Shay, Jackie E. and Stephens, Molly R. and Slatyer, Rachel A.",
   title = "Evolution of Ecological Niche Breadth", 
   journal= "Annual Review of Ecology, Evolution, and Systematics",
   year = "2017",
   volume = "48",
   number = "Volume 48, 2017",
   pages = "183-206",
   doi = "10.1146/annurev-ecolsys-110316-023003",
   url = "https://www.annualreviews.org/content/journals/10.1146/annurev-ecolsys-110316-023003",
   publisher = "Annual Reviews",
   issn = "1545-2069",
   type = "Journal Article",
  }

@article{Bebber2022,
    author = {Bebber, D. P. and Chaloner, T. M.},
    title = {Specialists, generalists and the shape of the ecological niche in fungi},
    journal = {The New phytologist},
    year = {2022},
    month = {02},
    volume = {234},
    number = {2},
    pages = {345--349},
    doi = {10.1111/nph.18005}
}

@article{Rossberg2010,
    author = {Rossberg, Axel G. and Farnsworth, Keith D. and Satoh, Keisuke and Pinnegar, John K.},
    title = {Universal power-law diet partitioning by marine fish and squid with surprising stability–diversity implications},
    journal = {Proceedings of the Royal Society B: Biological Sciences},
    volume = {278},
    number = {1712},
    pages = {1617-1625},
    year = {2010},
    month = {11},
    issn = {0962-8452},
    doi = {10.1098/rspb.2010.1483},
    url = {https://doi.org/10.1098/rspb.2010.1483},
    eprint = {https://royalsocietypublishing.org/rspb/article-pdf/278/1712/1617/566795/rspb.2010.1483.pdf},
}

@article{Hutchinson2022,
author = {Hutchinson, Matthew C. and Dobson, Andrew P. and Pringle, Robert M.},
title = {Dietary abundance distributions: Dominance and diversity in vertebrate diets},
journal = {Ecology Letters},
volume = {25},
number = {4},
pages = {992-1008},
doi = {10.1111/ele.13948},
url = {https://onlinelibrary.wiley.com/doi/abs/10.1111/ele.13948},
eprint = {https://onlinelibrary.wiley.com/doi/pdf/10.1111/ele.13948},
year = {2022}
}

@article{Qiao2016,
author = {Huijie Qiao  and Erin E. Saupe  and Jorge Soberón  and A. Townsend Peterson  and Corinne E. Myers  and David C. Collar  and Judith L. Bronstein },
title = {Impacts of Niche Breadth and Dispersal Ability on Macroevolutionary Patterns},
journal = {The American Naturalist},
volume = {188},
number = {2},
pages = {149-162},
year = {2016},
doi = {10.1086/687201},
URL = {https://www.journals.uchicago.edu/doi/abs/10.1086/687201},
eprint = {https://www.journals.uchicago.edu/doi/pdf/10.1086/687201}
}

@article{Lin2017,
author = {Lin, Long-Hui and Wiens, John J.},
title = {Comparing macroecological patterns across continents: evolution of climatic niche breadth in varanid lizards},
journal = {Ecography},
volume = {40},
number = {8},
pages = {960-970},
doi = {10.1111/ecog.02343},
url = {https://nsojournals.onlinelibrary.wiley.com/doi/abs/10.1111/ecog.02343},
eprint = {https://nsojournals.onlinelibrary.wiley.com/doi/pdf/10.1111/ecog.02343},
year = {2017}
}
\end{document}